\documentstyle[12pt,psfrag,epsfig]{article}
\newcommand{\bi}[2]{\vspace*{-2mm} \bibitem{#1}#2}

\newcommand{\ncm}[2]{\newcommand{#1}{#2}}

\ncm{\He}{\hbox{}{$^3$He}\hbox{ }}
\ncm{\Ht}{\hbox{}{$^3$H}\hbox{ }}

\ncm{\nmsp}{\!\!\!\!}      
\ncm{\nmspa}{\nmsp\nmsp}   
\ncm{\nmspb}{\nmspa\nmspa}   
\ncm{\nmspc}{\nmspa\nmspb}   
\ncm{\nmspd}{\nmspb\nmspb}   

\ncm{\noi} {\noindent}

\ncm{\OH} {\displaystyle {\frac{1}{2}}}
\ncm{\Oh} {\textstyle    {\frac{1}{2}}}
\ncm{\tH} {\textstyle    {\frac{3}{2}}}
\ncm{\oh} {\scriptstyle  {\frac{1}{2}}}
\ncm{\nh} {              {\frac{1}{2}}}

\newcommand{\dss}{\displaystyle}

\newcommand{\ap} { \hat p}
\newcommand{\app} { \hat {p'}}
\newcommand{\be} {\begin{equation}} \newcommand{\ee} {\end{equation}}
\newcommand{\ba} {$$\begin{array}}  \newcommand{\ea} {\end{array}$$}
\newcommand{\bea}{\begin{eqnarray}} \newcommand{\eea}{\end{eqnarray}}
\newcommand{\bl} {\begin{eqnarray}\begin{array}}
\newcommand{\el} {\end{array}\end{eqnarray}}

\ncm{\cP} {{\cal P}} \ncm{\cT} {{\cal T}}
\ncm{\phe}{\hphantom{=}}

\newcommand{\Ylp} {Y^*_{l'm'} \left(\hat{p}\;'\right)}
\newcommand{\Yl } {Y  _{l m } \left(\hat{p} \right)}

\newcommand{\JAtt} {\vec{j} \left( \vec{Q}/2+\vec{p}\;'-\vec{p},\:
                    \vec{Q}/2-\vec{p}\;'+\vec{p};\;2,3 \right)}
\newcommand{\JAttt} {\vec{j} \left( \vec{Q}/2+\vec{p_1}\;'-\vec{p_1},\:
                    \vec{Q}/2-\vec{p_1}\;'+\vec{p_1};\;2,3 \right)}

\newcommand{\vQ}  {\vec{Q}}
\newcommand{\vpp} {\vec{p}\;'}      \newcommand{\vp } {\vec{p}}
\newcommand{\vqp} {\vec{q}\;'}      \newcommand{\vq } {\vec{q}}
\newcommand{\vpw} {\vec{p}_2}       \newcommand{\vph} {\vec{p}_3}

\newcommand{\vpiw}{v_\pi (p_2)}     \newcommand{\vpih}{v_\pi (p_3)}
\newcommand{\vpik}{v_\pi (k)}

\newcommand{\vsw}{\vec\sigma(2)}    \newcommand{\vsh}{\vec\sigma(3)}
\newcommand{\spw}{\left( \vsw \cdot \vpw \right)}
\newcommand{\sph}{\left( \vsh \cdot \vph \right)}

\newcommand{\twh}{i \: {\left[ {\vec{\tau}(2)} \times {\vec{\tau}(3)}
                        \right] }_z}

\newcommand{\Cpi} {{\dss\frac{f^2_{\pi NN}}{2 \pi^2 \mpi}}}
\newcommand{\Fpi} {F_{\pi NN}}
\newcommand{\mpi} {m^2_\pi}
\newcommand{\Lpi} {\Lambda^2_\pi}

\begin{document}
\begin{center}
{\large\bf
 Partial Wave Decomposition \\
 for Meson Exchange Currents in Few-Nucleon Systems
}\vskip 5 mm
{\large                                               V.V.Kotlyar
}\vskip 5 mm
National Science Center, ``Kharkov Institute of Physics and Technology'' \\
Institute of Theoretical Physics \\
Kharkov 310108 Ukraine
 \vskip 5 mm

 {\large                H. Kamada\footnote{present address: Institut f\"ur
 Strahlen- und Kernphysik der Universit\"at Bonn,
 Nussallee 14-16, D-53115 Bonn, Germany} and  W. Gl\"ockle
 }\vskip 5 mm
 Institut f\"ur theoretische Physik II, \\
 Ruhr-Universit\"at Bochum, D-44780 Bochum, Germany
  \vskip 5 mm

  {\large                   J. Golak
  }\vskip 5 mm
  Institute of Physics, Jagellonian University, PL-30059 Cracow, Poland
  \end{center} \vskip 5 mm

\begin{abstract}

We develop an approach for calculating matrix elements of meson exchange currents
between 3N basis states in $(jJ)$--coupling and a 3N bound state.
The contribution generated by $\pi$-- and $\rho$-exchange are included in the
consideration.
The matrix elements are expressed in terms of multiple integrals in the momentum space.
We apply a technique of the partial wave decompositions
and carry out some angular integrations in closed form.
Different representations appropriate for numerical calculations
are derived for the matrix elements of interest.
The momentum dependences of the matrix elements are studied and
benchmark results are presented.
The approach developed is of interest for the investigations of deuteron--proton radiative
capture and \He photo-- and electrodisintegration when the interaction in the initial or
final nuclear states is taken into account by solving the Faddeev equations.
\end{abstract}

\renewcommand{\thesection}{\arabic{section}.}
\renewcommand{\thesubsection}{\arabic{section}.\arabic{subsection}}

\setcounter{section}0

    \def\theequation{\thesection \arabic{equation}}

\section{Introduction}\label{Introductn}
\setcounter{equation}0
\vskip -2ex

Various approaches have been developed to calculate the amplitudes of photo-- and
electrodisintegration of \Ht and \He and the radiative capture of nucleons by deuterons.
Methods employed and results obtained recently in studying these reactions can be found in
\cite{Meijgaard}--\cite{RoccoJoe}. Review article \cite{RoccoJoe} provides additional important references.

A common feature of investigations \cite{Meijgaard}--\cite{SandhasSchadow}
is the use of the solutions of the Faddeev--type equations for the 3N bound states.
While the results of \cite{Meijgaard} have been obtained
with the s--wave Malfliet--Tjon NN interaction,
realistic models of nuclear forces
(e.g., the Argonne, Bonn, Paris and Reid soft core potentials) have been
employed in \cite{Friar90}--\cite{SandhasSchadow}.

In
\cite{Meijgaard}, \cite{Friar90}, \cite{Fonseca}, \cite{FonsecaC53} and
\cite{IshiSasC45}--\cite{RoccoJoe}
the rescattering effects (the interactions in the initial or final states) are taken into
account in different ways.
For example, in \cite{IshiSasC45} the wavefunction (WF)
of the pd--system for the reaction
\be\label{RadCapTd} \vec{\hbox{d}}+\hbox{p} \to {^3\hbox{He}}+\gamma \ee
has been calculated partially in the coordinate and partially in the momentum
space representation applying the method of continued fractions,
whereas in \cite{Golak97} the initial state interaction for the same reaction
has been considered
applying an iterative procedure to a Faddeev--like equation in the momentum space and
summing up the resulting multiple scattering series via the Pad\'e method.

Calculations \cite{FonsecaC53} and \cite{SandhasSchadow} are based
on separable approximation \cite{Graz85},\cite{Graz86}
constructed with the help of the Ernst--Shakin--Thaler method\cite{EST}.
The use of correlated hyperspherical harmonic WFs for the 3N systems is mentioned in \cite{RoccoJoe}.

An essential element of all these treatments is a model of the electromagnetic (EM)
current for the 3N system. In
\cite{Meijgaard},  \cite{Fonseca}--\cite{SandhasSchadow}
the current is chosen following the basic idea of the impulse approximation, i.e., it is
assumed that the current is an one--body operator.
Two--body contributions to the current (interaction or meson exchange currents
(MEC)) have been included either implicitly
\cite{Meijgaard}, \cite{Fonseca}--\cite{SandhasSchadow}
as pointed out below or in an explicit form
\cite{Friar90}--\cite{Spin96},\cite{RoccoJoe}.
Earlier the contributions of MEC to the magnetic multipoles have been
considered in investigations \cite{Torre83}--\cite{Jourdan86}
of the radiative capture (\ref{RadCapTd}).

In this connection, it is pertinent to recall
(see, e.g., \cite{MesonNucl},\cite{Ciofi})
that the necessity of the inclusion of MEC results from the continuity equation (CE)
\be\label{CEJx}
\partial _\mu j^\mu (x) =0
\ee
or its equivalent
\be\label{CEJ0}
\left[H,j^0(0)\right]=\left[\vec P, {\vec j}(0)\right], \ee
where
$j^{\mu}(\vec x)=\left(j^0(\vec x),{\vec j}(\vec x)\right)$
is the current density operator and $\vec P$ is the total momentum operator for a given
system (nucleus).
The nuclear Hamiltonian $H=K+V$ consists of the kinetic energy part $K$ and the
interaction part $V$. The latter is isospin dependent,
and, in general, can be
nonlocal, i.e., dependent not only on the nucleon coordinates but also on
their derivatives.

For a system with such a  Hamiltonian where the forces can depend
arbitrarily  on  nucleon velocities a conserved EM current
can be constructed following, for example, a prescription \cite{Korchin84}.

As is well known (see, e.g., \cite{Kazes},\cite{Shebeko87}),
the CE for the current operator is not a sufficient condition for ensuring  gauge
independence (GI) of the reaction amplitudes.
Indeed, the GI requirement reads
\be\label{GIreq} q^\mu\left<f\right|j_\mu (0)\left|i\right>=0, \ee
where $q^\mu$ is the four momentum transfer. Eq{.}(\ref{GIreq}) is fulfilled if
the operator $j_\mu (0)$ meets Eq{.}(\ref{CEJ0}) and both the initial
$\left|i\right>$ and final
$\left|f\right>$ nuclear states are exact eigenvectors of $H$.
The implementation of these  requirements is a quite demanding task in practical
calculations because one has to take care of the definite consistency
between the current and the Hamiltonian and
``good'' WFs are needed both for initial and final states. In this respect, an exception
is the calculation of the charge form factor (FF) for a nucleus (for instance \He). In this case
one can show that the one--body $(j^{[1]}(\vec x))$
and many--body meson $(j^{[{meson}]}(\vec x))$
contributions to the decomposition
$j(\vec x)=j^{[1]}(\vec x)+j^{[{meson}]}(\vec x)$
satisfy Eq.(\ref{GIreq}) separately
(see Refs{.}\cite{Kamada92}--\cite{Burov96}).
Of course, it does not mean that one can avoid the inclusion of MEC.

For the nuclear processes with real or virtual photons Eq.(\ref{GIreq})
is useful for an effective account for MEC effects.
Using Eq.(\ref{GIreq}) the matrix elements of the longitudinal component of the EM current
can be expressed through the corresponding matrix element of its time component.
This replacement makes sense due to the fact that the latter is subject to the meson exchange
influence to a lesser extent than the space component of the current (cf. classification of
the MEC given by Friar \cite{Friar83} and discussion in Sect. 2).
This trick is often employed in the treatments of inelastic electron scattering on nuclei
in order to express the differential cross sections in terms of
the respective structure functions
\cite{Meijgaard},\cite{d3He}--\cite{Spin96},
\cite{IshikawaA107}--\cite{IshikawaC57}, \cite{deForest}.

One can also write the amplitudes in a GI form via  extended Siegert theorem
\cite{FriarFallieros}--\cite{ShebLevch93}
expressing them through
the strengths of the electric and magnetic fields and the generalized electric and magnetic
dipole moments of a system (nucleus)

The Siegert theorem  is widely used as a recipe to incorporate
a part of MEC effects within the long wavelength approximation.
However, all the applications of the Siegert theorem imply that the initial
and final states involved in the transition matrix elements are solutions of the
Schr\"odinger equation with the same Hamiltonian.
This important point has been violated in some previous calculations
(see, e.g., \cite{KlepackiA550} and \cite{AuflegerA364}).

The initial and final nuclear states have been treated consistently in studies
\cite{Meijgaard}, \cite{Friar90}, \cite{Fonseca}--\cite{FonsecaC53} and
\cite{IshiSasC45}--\cite{Jourdan86}
of inelastic  electron scattering on 3N nuclei and the
radiative capture.
The methods developed in three--particle scattering theory
\cite{Gloeckle83}--\cite{GloecklePhysRep}
provide a rigorous description of the nuclear states belonging to the continuum.
All relevant components of  realistic NN potentials can be included. In
\cite{IshikawaA107}--\cite{Golak97} rescatterings in the two-- and three--body channels for
the \He breakup by electrons are treated on the same footing
by solving the Faddeev--like integral equation
\be\label{FlEqU}
\left|U_\kappa\right>=tG_0(1+P)j_\kappa (\vec Q)\left| \Psi _{bound}\right>
+tG_0P \left|U_\kappa\right>,
\ee
where $j_\kappa (\vec Q) \: (\kappa=0,\pm1)$ are the spherical components of
the Fourier transform of the current ope\-ra\-tor.
The calculations have been performed in the $pq\alpha$--representation,
where p and q are the Jacobi momenta and $\alpha$ stands for discrete quantum
numbers in $(jJ)$-- or $(LS)$--coupling. As is noted in \cite{GolakC52}, the block
$j^\mu(\vec Q)\left|\Psi _{bound} \right>$
is the only essentially new structure in comparison to the nucleon--deuteron
scattering theory.

Only the one--nucleon part of the EM current has been utilized in
\cite{IshikawaA107}--\cite{Golak97}.
Unlike this, in studies \cite{KShS}--\cite{Spin96}, \cite{KSh87} of the two--body \He
disintegration by real and virtual photons at the intermediate energies  special attention
has been paid to the MEC contributions with no final state interaction included. The
two--body currents \cite{Riska11,Mathiot173} due to the $\pi$-- and $\rho$--meson
exchanges and MEC with the excitation of the $\Delta$--isobar in intermediate states have
been taken into account.
In \cite{KShS}--\cite{Spin96}, \cite{KSh87}
the reaction amplitudes have been evaluated
with a convenient parameterization
\cite{HGS} for the Faddeev components $\Psi^{(1)}$ of the \He WF. This
component has been computed with the Reid soft core potentials in \cite{HajdukSauer79}.

The simultaneous consideration of the MEC and rescattering effects in the theory of
the reactions still remains to be carried out.
The purpose of the present paper is to develop
a method for the calculation of the matrix elements
%
%
of the two body current between the 3N basis states and the 3N bound state.
A particular emphasis is laid on a compatibility with the approach implemented in
\cite{IshikawaA107}--\cite{Golak97}.
We treat the matrix elements in the momentum representation and use the basis 
in the space of 3N states formed by the vectors with the quantum numbers
in $(jJ)$--coupling.


We aim to elaborate a technique which can allow us to provide the
GI of the reaction amplitudes in the calculations with the WFs for
realistic models of nuclear forces.

This paper is organized as follows.
In Section 2 the transition matrix elements of the two--body currents are transformed to forms which are
suitable for numerical calculations.
They are expressed in terms of two- and four--dimensional angular integrals.
Section 3 and Appendices A--C are devoted to the reduction of the angular integrals. The partial wave
decomposition of the $\rho$--meson exchange currents is given in Appendix D.
We show two ways how one can analytically carry out three integrations over the angles.
The explicit expressions for the matrix elements of interest are presented.
Section 4 gives benchmark results in the case of the $\pi$--meson exchange currents. We
use numerical methods to 
 evaluate angular integrals employing MEC matrix between the 3N basis states 
and compare the results with ones
obtained within the framework of the approaches proposed above.
Section 5 contains the summary and  a discussion.
The Appendices contain detailed expressions for the matrix elements and the integrals
involved. In particular, in Appendix D some angular integrals appearing in the derivation are
calculated in a closed form.


%
\section{\rm Contributions of Two-Body Currents }\label{Transform}
\setcounter{equation}0
\vskip -2ex

Amplitudes of the reactions
\ba{lll}
\gamma+{^3\mbox{H}}\mbox{e} \to \mbox{p}+\mbox{d},&
&\mbox{e}+{^3\mbox{H}}\mbox{e} \to \mbox{e}'+\mbox{p}+\mbox{d}, \nonumber\\
\gamma+{^3\mbox{H}}\mbox{e} \to \mbox{p}+\mbox{p}+\mbox{n},&
&\mbox{e}+{^3\mbox{H}}\mbox{e} \to \mbox{e}'+\mbox{p}+\mbox{p}+\mbox{n} \nonumber
\ea
can be written as
\be\label{Nt}
N^\mu=\left\langle \Psi_f^{(-)}
\left|\vphantom{\Psi_f^{(-)}}
j^\mu(\vQ) \right|\vphantom{\Psi_f^{(-)}}
\Psi_{i} \right\rangle,
\qquad (\mu=0,1,2,3),
\ee
where $j^\mu ( \vQ )$ is the Fourier transform of the current density $j^\mu(\vec x\:),$
the initial $\left|\Psi_{i}\right\rangle$
and final   $\left|\Psi_f^{(-)}\right\rangle$
states are eigenvectors of one and the same Hamiltonian $H$.

Our consideration refers to the case where the current
$ j(\vQ)=j^{[1]}(\vQ)+j^{[2]}(\vQ)$, i.e., consists of a
one--body current
\be j^{[1]}(\vQ)=j(\vQ;1)+j(\vQ;2)+j(\vQ;3) \ee
and a two--body current
\be j^{[2]}(\vQ)=j(\vQ;1,2)+j(\vQ;1,3)+j(\vQ;2,3).\ee
The operators $j^{[1]}(\vQ;a)$ and $j^{[2]}(\vQ;b,c)$ depend on the variables
of the nucleons with the labels $a$ and $b,c$, respectively.
Taking into account these decompositions and the permutation properties
of the states, one can write
\be\label{N123}
N^\mu=3 \left\langle \Psi_f^{(-)}
\left|\vphantom{\Psi_f^{(-)}} j^\mu(\vQ;1) + j^\mu(\vQ;2,3)
\right|\vphantom{\Psi_f^{(-)}} \Psi_{i} \right\rangle.
\ee

The method for calculation of the one--body contributions to
amplitude (\ref{N123}) has been worked out earlier and used in studying
the elastic electron scattering off \He \cite{Kamada92},
the \He disintegration by electrons \cite{IshikawaA107}--\cite{IshikawaC57} and
the proton--deuteron radiative capture \cite{Golak97}.
The detailed description of the corresponding formalism can be found in the above articles.
Here we concentrate on a treatment of the two--body contributions.

In the contemporary theories of EM interactions with nuclei, where
many--body currents are generated by meson exchange, the time
component $j^{[2]}_0$ is of order  $v^2/c^{\:2}$ compared to
$j^{[1]}_0$ which is of order $(v/c)^0$,
while ${\vec j}^{\:[2]}\sim{\vec j}^{\:[1]} = O( v/c) $.
Therefore, neglecting the $v^2/c^{\:2}$--contributions, we assume
\be j_0 \simeq j^{[1]}_0 \ee and
\be \vec{j} = \vec{j}^{\;[1]} +  \vec{j}^{\;[2]}. \ee
In other words, the dominant MEC contribution to the reaction amplitude
originates from the space part of the current,
$
\left\langle \Psi_f^{(-)}  \left| \vec{j}(\vQ;2,3) \right|
\Psi_{i} \right\rangle.
$

To get the matrix elements of interest
let us project the vector $j^\mu(\vQ;2,3)\left|\Psi_{i}\right>$ on the basis
\be\label{pqalpha}
\left| pq\alpha\right> \equiv
\left| pq \: (ls)j \: (\lambda\Oh)I \: (jI)JM; \: (t\Oh)TM_T\right>_1
\ee
used throughout Refs{.}\cite{IshikawaA107}--\cite{Golak97}, \cite{Kamada92},
\cite{Gloeckle83}--\cite{GloecklePhysRep}.

\begin{figure}[hbt]
\centerline{
\epsfysize=57mm
\psfig{file=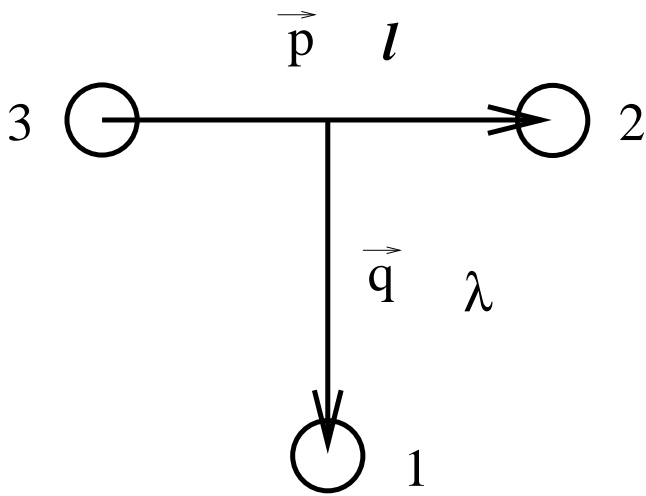,scale=0.7}
}
\caption{Jacobi momenta.}
\label{fig0}
\end{figure}

In Eq. (\ref{pqalpha}) the subscript ``1'' is to indicate the definite choice
of the Jacobi momenta
\be \label{jacobi1} \label{jacobi2}
\vec p = {1 \over 2} (\vec { k_2 } - \vec { k_3 } ), \qquad
\vec q = {1 \over 3} ( 2 \vec k_1 - \vec k_2 -\vec k_3 )
\ee
and the angular momentum coupling scheme as it is depicted in Fig{.}\ref{fig0}.
The nucleon momenta are denoted by $\vec k_a \; (a=1,2,3)$.
The orbital angular momentum, the total spin, and the angular momentum
in the two--body subsystem 2--3 are $l,s$ and $j,$ respectively.
The orbital angular momentum $\lambda$ of nucleon 1 and its spin couple to
the angular momentum $I$ of this particle.
The angular momentum of the 3N system and its magnetic quantum number are denoted by
$J$ and $M$.
The total isospin of particles 2 and 3 is $t$, while $T,M_T$ refer to
the isospin of the 3N state and its projection.
The set of states (\ref{pqalpha}) is orthonormalized and complete
(see, e.g. \cite{GloecklePhysRep}, Section III).
Since the nuclear states
$\left|\Psi_f^{(-)}\right>$ and $\left|\Psi_{i} \right>$
are antisymmetric
we can consider the subset of vectors (\ref{pqalpha}) constrained by requirement
$$ l+s+t=2n+1, \qquad (n=0,1,2,\ldots). $$

It is convenient to split the whole array of discrete quantum numbers $\alpha$,
separating the isospin part of vectors (\ref{pqalpha})
\be\label{pqgamma}
\left| pq\alpha\right> =
\left| pq \alpha_J \right>\left| \alpha_T  \right>,
\ee
where 
\be
\left| \alpha_J \right> = \left| (ls)j,(\lambda {1\over2})I, (jI)J M \right> 
\ee
and 
\be
\left| \alpha_T \right> = \left| (t {1 \over 2}) T M_T \right>
\ee

The matrix elements of the two--body currents we shall consider have the form
\[
\left\langle \vec{k}'_2 \vec{k}'_3 \left|
\vec j \left( \vQ;2,3 \right)
\right| \vec{k}_2 \vec{k}_3 \right\rangle     \equiv 
\delta\left(\vec{k}'_2 + \vec{k}'_3
-\vQ - \vec{k}_2 - \vec{k}_3 \right)
\vec j \left(
\vec{k}'_2 - \vec{k}_2, \vec{k}'_3 - \vec{k}_3; \; 2,3  \right),
\]
\begin{eqnarray}\label{J23me}
\end{eqnarray}
or
in 3-particle space (without the overall momentum conserving delta-function)
\bl{rl}\label{J23pq}
{\left\langle \vpp \vqp 
{\left| \vec j {\left( \vQ ;2,3 \right)} \right|}
\vp \vq  \right\rangle} = &
\delta\left(\vqp - \vq + \vQ /3 \right) \\ &  \times \JAtt.
\\ & = \delta\left( \vec p_2 + \vec p_3 -\vQ   \right)
\vec j \left( \vec p_2 , \vec p_3 ; 2,3 \right)
\el
with 
\bl{rl}
&
\vec p_2 \equiv \vec {k'_2} - \vec k_2 = \vQ /2 + \vec {p'} -\vec p
\\ & 
\vec p_3 \equiv \vec {k'_3} - \vec k_3 = \vQ /2 - \vec {p'} + \vec p
\label{p23}
\el
The corresponding matrix elements of the one-body current are 
\be
\left\langle \vec {k'_1} \left| \vec j (\vec Q ; 1 ) 
\right| \vec { k_1 } \right\rangle 
= \delta ( \vec {k_1 ' } - \vec { k_1 } -\vec Q ) 
\vec j ( \vec { k_1'} , \vec Q ; 1 ) 
\ee
or in 3-particle space and in the lab frame
\be\label{J1pq}
\left\langle \vec {p'} \vec {q'} \left| 
\vec j ( \vec Q ; 1 ) \right| \vec p \vec q \right\rangle 
= \delta ( \vec p - \vec {p'} ) \delta ( \vec {q '} - \vec 
q -{ 2 \over 3 } \vec Q ) 
\vec j ( \vec {q'} + { 1 \over 3 } \vec Q , \vec Q ; 1 ) 
\ee
Eqs.(\ref{J23pq}) and (\ref{J1pq}) exhibit the important shift 
prescriptions 
$\delta\left(\vqp - \vq + \vQ /3 \right) $ and 
$ \delta ( \vec {q '} - \vec
q -{ 2 \over 3 } \vec Q ) $.
The one-body current matrix element has been dealt with in \cite{Kamada92,GolakC52} in
great detail and therefore we shall concentrate now only on (\ref{J23pq}).

Because the factor $\JAtt$ in (\ref{J23pq}) has no dependence on 
$\vec q$ we can treat the $\vec p-$ and $\vec q-$ 
spaces separately. 
We regard  the partial wave decomposition of 
the two-body current formed in terms of the basis states (\ref{pqalpha}) or 
(\ref{pqgamma}).

\[ 
\left\langle  p' q' \alpha ' \left| \vec j ( \vec Q ; 2,3 ) \right|
p q \alpha \right\rangle 
\]
\[
 =  \int d\vec {p'_1} \int d \vec {q'_1}
 \int d\vec {p_1} \int d\vec {q_1} 
\left\langle p' q' \alpha \vert \vec {p'_1} \vec {q'_1 } 
\right\rangle 
\left\langle \vec {p'_1 } \vec { q'_1} \left| j ( \vec Q ; 2,3 ) 
\right| \vec {p_1} \vec {q_1} \right\rangle 
 \left\langle \vec { p_1 } \vec { q_1 } \vert p q \alpha
\right\rangle
\]
\be
 =  \int d\vec {p'_1} \int d \vec {q'_1}
 \int d\vec {p_1} \int d\vec {q_1}
\sum _{ \mu ' \mu } C(j' I' J' , \mu ', M' -\mu')
C(j,I,J,\mu , M-\mu) 
\ee
\[
\times
 {\cal Y}_{j' \mu'} ^* 
( \hat {p'_1}) { {\delta ( p'_1 - p' ) } \over {p'_1}^2 } 
{\cal Y } _{I', M'-\mu '} ^* ( \hat {q'_1} ) 
{{\delta ( q'_1 -q') } \over { q '_1 } ^2 } 
\delta ( \vec {q_1} -\vec {q'_1} - {1 \over 3 } \vec Q ) 
\]
\[  \times
\left\langle \alpha _{T'} \left|  
\JAttt \right| \alpha_T \right\rangle 
\]
\[  \times
{\cal Y}_{j \mu}  
( \hat {p_1}) 
{\cal Y } _{I, M-\mu} ( \hat {q_1} ) 
{ {\delta ( p_1 - p ) } \over {p_1}^2 } 
{{\delta ( q_1 -q) } \over { q_1 } ^2 }
\]
The momentum states $\vert \vec p_1 \vec q_1 \rangle $ and $ \vert
\vec {p '} _1 \vec {q '} _1 \rangle $ are formed out of Jacobi momenta of the
type  (\ref{jacobi1}). We obtain

\[
\left\langle  p' q' \alpha ' \left| \vec j ( \vec Q ; 2,3 ) \right|
p q \alpha \right\rangle
\]
\[
 =
\sum _{ \mu ' \mu } C(j' I' J' , \mu ' , M' -\mu')
C(j,I,J,\mu , M-\mu)
\]
\[  \times
\left( \int d\hat  {p'_1}\int d\hat {p_1}  {\cal Y}_{j' \mu'} ^*  
( \hat {p'_1})  \left\langle \alpha _{T'} \left| 
  \vec j \left( \vec Q/2 + p'  \hat  p'_1 - p\hat p_1, \vec Q /2 - 
p' \hat p' _1 +
p \hat  p_1 ; 2,3  \right)  \right| \alpha_T \right\rangle
{\cal Y}_{j \mu}  
( \hat {p_1}) \right)
\]
\[  \times
\left( \int d\hat {q'_1} {\cal Y } _{I', M'-\mu '} ^* ( \hat {q'_1} )
{{\delta ( q -\vert q' \hat {q'_1} + {1 \over 3} \vec Q \vert ) } \over { q } ^2 }
{\cal Y } _{I, M-\mu}  ( \widehat { q'  \hat {q'_1} + {1 \over 3} \vec Q} ) 
\right)
\]
\[
= 
\sum _{ \mu ' \mu } C(j' I' J' , \mu' , M' -\mu')
C(j,I,J,\mu , M-\mu)
\]
\[
\times 
\vec I_2 ( p',p,Q; (l's')j' \mu ' \alpha_{T'},(ls)j \mu \alpha_T ) 
\]
\[  \times
I _3( q',q,Q ; (\lambda' {1\over2}) I' M'-\mu ' ,( \lambda {1 \over
2}) I M- \mu )
\]
with 
\be\label{EQI2} 
\vec 
I_2 (  p',p,Q; (l's')j' \mu ' \alpha_{T'},(ls)j \mu, \alpha_T )
= 
\ee\[
\int d\hat  {p'}\int d\hat {p}  {\cal Y}_{j' \mu'} ^*
( \hat {p'})  \left\langle \alpha _{T'} \left| 
 \JAtt  \right| \alpha_T \right\rangle  {\cal Y}_{j \mu}
( \hat {p})
\]
and 
\be \label{EQI3} 
I _3( q',q,Q ; (\lambda' {1\over2}) I' M'-\mu ' ,( \lambda {1 \over
2}) I M- \mu ) 
\ee
\[
= 
 \int d\hat {q'} {\cal Y } _{I', M'-\mu '} ^* ( \hat {q'} )
{{\delta ( q -\vert \vec {q'} + {1 \over 3} \vec Q \vert ) } \over { q } ^2 }
{\cal Y } _{I, M-\mu}  ( \widehat {\vec {q'} + {1 \over 3} \vec Q} )
\]
Finally $ {\cal Y}_{j\mu} (\hat p)$ and ${\cal Y}_{I,\nu} (\hat q)$ are 
defined as
\be 
{\cal Y} _{j \mu} (\hat p ) \equiv \sum _m C (l s j , \mu -m , m, \mu ) 
Y_{l,\mu -m} (\hat p) \vert s m \rangle
\ee
and
\be 
{\cal Y} _{I \nu} (\hat q ) \equiv  \sum _m C (\lambda {1\over2} I , \nu -m ,
 m, \nu  ) 
Y_{\lambda ,\nu -m} (\hat q) \vert {1\over2} m \rangle {\rm .}
\ee
Now we use the result (\ref{EQI3}) to evaluate the decisive 
building block $
\left\langle 
p q \alpha \left| \vec j ( \vec Q; 2,3 ) \right| \Psi_{bound} \right\rangle $.
We obtain
\be
 \left\langle p' q'  \alpha ' \left| \vec j ( \vec Q ; 2,3 ) 
\right| \Psi_{bound} 
\right\rangle
= \sum _{ \alpha }
\int p^2 d p  q^2 d q  
 \left\langle p' q'  \alpha ' \left| \vec j ( \vec Q ; 2,3 )
 \right| p q \alpha \right\rangle 
\left\langle p q \alpha | \Psi_{bound} \right\rangle 
\ee
\[
 = \sum _{ \alpha } \int p^2 d p  \sum_{\mu \mu ' } 
C(j' I' J' , \mu , M' -\mu' )
C(j,I,J,\mu , M-\mu )
\]
\[
\times  
\vec I_2 ( p',p,Q; (l's')j' \mu ' \alpha _{T'},
(ls)j \mu \alpha_T ) 
\]\[  \times
\tilde I _3( p, q',Q ; (\lambda' {1\over2}) I' M'-\mu ' ,( \lambda {1 \over
2}) I M- \mu )
\]
with  $\tilde I_3$ given as 
\be\label{EQI3PSI}
 \tilde I _3(p, q',Q ; (\lambda' {1\over2}) I' M'-\mu ' ,( \lambda {1 \over
2}) I M- \mu ) = 
\ee
\[
 \int d\hat {q'} {\cal Y } _{I', M'-\mu '} ^* ( \hat {q'} )
{{ \left\langle p, \vert  \vec {q'} + {1 \over 3} \vec Q \vert \alpha \vert
 \Psi_{bound} \right\rangle  }  }
{\cal Y } _{I, M-\mu}  ( \widehat {\vec {q'} + {1 \over 3} \vec Q} ){\rm .}
\]
The angular integration over $\hat {q'}$ can easily be  performed as demonstrated 
in Appendix A.
The integral $I_2$ is more complicated because the operator $j$ 
also acts on the spin space. To proceed  
we need some specific  expressions of the two-body currents.

Let us return to integrals (\ref{EQI2}) and examine the spin--isospin
structure of the integrands. In this section we consider
model \cite{Riska11},\cite{Mathiot173} of $\pi$--meson exchange currents
and assume that
\be\label{Jsgpion}
\vec{j}^{\:[2]}=\vec{j}^{seagull} + \vec{j}^{pionic}.
\ee
The corresponding contributions to $\vec j {\left(\vpw,\vph ;2,3 \right)}$ read
\be\begin{array}{lll}\label{Jseag}
\lefteqn{
{\vec{j}^{seagull} \left( \vpw, \vph;\;2,3 \right)}
} &&\\&=&
F_1^V\Bigl(\vpiw \: \spw \: \vsh - \vpih \: \vsw \: \sph \; \Bigl)
\\& \times & {\twh},
\end{array}\ee
\be\begin{array}{lll}\label{Jpion}
\lefteqn{
{\vec{j}^{pionic} \left(\vpw, \vph;\;2,3 \right)}
} &&\\&=&
F_1^V\left( \vpw - \vph \right) \: \spw \: \sph \:
{\dss{{\vpih - \vpiw} \over { p_2 ^2 -p_3 ^2 }}}
\\& \times &{\twh},
\end{array}\ee
where
\be\label{vpi}
\vpik=
 \Cpi \: {\dss \frac{1}{\mpi + k^2}} \: {\Fpi^2(k^2)},
\ee
further $F_1^V=F_1^p-F_1^n$ is the isovector nucleon form factor,
$f_{\pi NN}$ is the pseudovector $\pi N$ coupling constant,
$m_\pi$ is the pion mass and
${\Fpi(k^2)}$ is the strong $\pi NN$ form factor.

The operator $\vec{j}\left(\vp_2,\vp_3;2,3\right)$ can be written as
\be\begin{array}{lll}\label{MECSpIsoSp}
\lefteqn{
{\vec{j}\left(\vp_2,\vp_3;2,3\right)}
} &&\\&=&
F_1^V \; \sum\limits_{k\kappa}
{\vec O}^{k\kappa}\left(\vp_2,\vp_3\right) \;
\left\{ \sigma(2) \otimes \sigma(3) \right\}_{k\kappa}
\;{\twh},
\end{array}\ee
where $\left\{ \sigma(2) \otimes \sigma(3) \right\}_{k\kappa}$ is
the irreducible tensor product of rank $k$ ($k$=0,1 and 2).

According to (\ref{Jsgpion})--(\ref{Jpion})
the functions ${\vec O}^{k\kappa}\left(\vp_2,\vp_3\right)$ are
\be\label{Osgpn}
{\vec O}^{k\kappa}={\vec O}^{k\kappa}_{seagull}+{\vec O}^{k\kappa}_{pionic},
\ee where
\be\label{Osgkk}\begin{array}{lll}
\lefteqn{
\left[ { O}^{k\kappa}_{seagull}\left(\vp_2,\vp_3\right) \right]_\zeta
} &&\\&&
={\dss\sum\limits_\xi} C ( 1 1 k, \xi\zeta\kappa)
(-1)^\xi
\left[ \vpiw (p_2)_{-\xi} + (-1)^{k+1}\vpih (p_3)_{-\xi} \right]
\end{array}\ee
with $\zeta,\xi=0,\pm1$ and
\be\label{Opnkk}\begin{array}{lll}
\lefteqn{ 
{\vec O}^{k\kappa}_{pionic}( \vpw ,\vph )
=\left(\vpw-\vph\right)
\left( p_2^2 - p_3 ^2  \right)^{-1}
\left( \vpih-\vpiw \right)
\left\{ p_2\otimes p_3 \right\}^{k\kappa}
}&&\\&&
={ { \vpw-\vph } \over {  p_2^2 - p_3 ^2  } 
} \left( \vpih-\vpiw \right) \sum _{\xi} 
C(1 1 k, \xi, \kappa-\xi, \kappa)(-)^\kappa
(p_2)_{-\xi}(p_3)_{\xi-\kappa} 
\end{array}\ee
Inserting (\ref{MECSpIsoSp}) into (\ref{EQI2}) we get
\be\label{EQI22}\begin{array}{lll}
\lefteqn{
 \vec I_2(p,p',Q; (l's') j' \mu ' \alpha _{T'} , (l s) j \mu  \alpha_T )
} &&\\&&
=  F_1^V {\dss\sum_{m'_l m'_s m_l m_s}} C (l's'j',m'_l m'_s \mu') C (lsj, m_l m_s \mu)
\\&&
\times {\dss\sum_{k\kappa}}
\vec M^{l'm'k\kappa}_{l m}(p',p; \vQ)
\left\langle s' m'_s \left|
\left\{ \sigma(2)\otimes\sigma(3) \right\}_{k\kappa} \right|sm_s \right\rangle
\\&&
\times \left\langle \alpha_T' \left|
\:i\:[ \tau (2) \times \tau (3)]_z
\right|\alpha_T \right\rangle.
\end{array}\ee
The angular integrals
\be\label{EQ4F}\begin{array}{lll}
\lefteqn{
\vec M^{l'm'k\kappa}_{l m}(p',p; \vQ)
} &&\\&&
={\dss\int d\hat p \int d \hat{p}'} \; Y _{ l'm' }^* (\hat{p}')
{ \vec O} ^{ k \kappa }
\left(\vQ/2+\vpp-\vp,\: \vQ/2-\vpp+\vp\right)
Y_{l m} ( \hat p )
\end{array}\ee
are analyzed in the subsequent sections.

The spin and isospin matrix elements have the form
\be\label{SpinPART}
\left\langle s' m' \left|
\left\{ \sigma (2) \otimes \sigma (3) \right\}_{k\kappa}
\right| s m \right\rangle
= 6 \sqrt{ \hat{s}\hat {s}' } C( k s s',\kappa m m')
\left\{ \begin{array}{ccc}
\oh & \oh & s ' \\ \oh & \oh & s \\ 1 & 1 & k
\end{array} \right\}{\rm ,}
\ee
\be\label{ISOPART}\begin{array}{lll}
\lefteqn{
\left\langle \alpha'_T \left|
 i [ {\vec \tau}(2) \times {\vec \tau} (3) ]_z
\right| \alpha_ T \right\rangle
} &&\\&=&
  12 \sqrt{3}
(-1)^{ 1 + t + \frac12 + T'} \sqrt{\hat{t}\:\hat{t'}} \,
C(1 \frac12 T'; 0 M_T M_T )
\\&\times&
                 \left \{ { \begin{array}{ccc}
                       1  &  t   & t' \\
                       \frac12 & T' & \frac12
                   \end{array}} \right \}
           \;  { \left \{ { \begin{array}{ccc}
                       1  &  1   & 1 \\
                       \frac12  &  \frac12   & t \\
                       \frac12  &  \frac12   & t'
                   \end{array}} \right \} },
\vphantom{\begin{array}{c} M\\M\\M\\M \end{array}}
\end{array}\ee
where $\hat a = 2a +1 $.
It has been used in Eq{.}(\ref{ISOPART}) that
the isospin of the initial state $T=1/2$.

Below we develop and study the potentialities of diverse approaches for
evaluation of the angular integrals (\ref{EQ4F}). Two different
method for further reduction of the matrix elements
are presented in Section~3 and  Appendices~B and C.
While Section 3 and Appendix B deal with the $\pi$--meson exchange currents,
the partial wave decomposition of the $\rho$--meson contributions
\cite{Riska11},\cite{Mathiot173} is given in Appendix~C.

Both approaches based on the analytical methods end up with 
two--dimensional angular integrations. These integrals and
the four--fold ones in (\ref{EQ4F}) ( called 4FI in the following ) have been calculated numerically.

%
\section{Reduction of the Multiple Integrals }\label{ReductMI}
\setcounter{equation}0
\vskip -2ex

In this section we show how one can reduce the dimensionality of
the angular integrals (\ref{EQ4F}) and transform them to the form appropriate for the
numerical evaluation.
It is convenient to consider the cyclic components
\be\label{M4Cyclic}
M^{l'm'k\kappa}_{lm\mu}(p',p;\vQ)
={\dss\int d\ap \int d\app} \; \Ylp O^{k\kappa}_\mu
\left(\vpw,\vph
\right) \Yl
\ee
of tensors (\ref{EQ4F}). We denote
\be\label{vecpn}
\vec{p}_n=\Oh\vQ+(-1)^n (\vpp-\vp), \qquad (n=2,3).
\ee

As it follows from Eq.(\ref{Osgpn})
\bea
M^{l'm'k\kappa}_{lm\mu}(p',p;\vQ) &=&
M^{l'm'k\kappa}_{lm\mu}(p',p;\vQ;\:seagull)
\nonumber\\ &+&
M^{l'm'k\kappa}_{lm\mu}(p',p;\vQ;\:pionic).
\eea

From Eqs.(\ref{Osgkk}),(\ref{Opnkk}) we get
\be\label{M4seagull}
\begin{array}{lll}\lefteqn{
M^{l'm'k\kappa}_{lm\mu}(p',p;\vQ;\:seagull) =
(-1)^{k+\kappa+\mu} {\dss\sum_\nu} (11k,\nu,-\mu,-\kappa)
} &&\\&\times&
\left(\Oh Q_\nu\left(
G^{l'm'}_{lm}(2)+(-1)^{k+1} G^{l'm'}_{lm}(3)
\right) +
G^{l'm'}_{lm\nu}(2)+(-1)^{k} G^{l'm'}_{lm\nu}(3)
\right)
\end{array}\ee
and
\be\label{M4pionic}                                
\begin{array}{lll}\lefteqn{
M^{l'm'k\kappa}_{lm\mu}(p',p;\vQ;\:pionic)=
(-1)^{k+\kappa} {\dss\sum_{\kappa_2\kappa_3}}
(11k,\kappa_2\kappa_3,-\kappa)
} &&\\&\times&
\left(
 \Oh Q_{\kappa_2} Q_{\kappa_3} F^{l'm'}_{lm\mu}
-Q_{\kappa_2}  F^{l'm'}_{lm\mu\kappa_3}
+Q_{\kappa_3}  F^{l'm'}_{lm\mu\kappa_2}
-              2 F^{l'm'}_{lm\mu\kappa_2\kappa_3}
\right),
\end{array}\ee
where                                              
\bea
G^{l'm'}_{lm}(n)   &=& {\dss\int d\ap \int d\app} \; \Ylp\Yl v(p_n),
\label{Glm}\\
G^{l'm'}_{lm\mu}(n)&=& {\dss\int d\ap \int d\app} \; \Ylp d_\mu\Yl v(p_n),
\label{Glmu}\\
&&(n=2,3)\nonumber
\eea and \bea
F^{l'm'}_{lm\mu_1}
&=& {\dss\int d\ap \int d\app} \Ylp d_{\mu_1} \Yl                     H(p_2,p_3),
\label{Fmu1}\\
F^{l'm'}_{lm\mu_1\mu_2}
&=& {\dss\int d\ap \int d\app} \Ylp d_{\mu_1}d_{\mu_2} \Yl            H(p_2,p_3),
\label{Fmu2}\\
F^{l'm'}_{lm\mu_1\mu_2\mu_3}
&=& {\dss\int d\ap \int d\app} \Ylp d_{\mu_1}d_{\mu_2}d_{\mu_3} \Yl   H(p_2,p_3).
\label{Fmu3}\eea
The explicit expressions for the scalar functions $v(p_n)$ and $H(p_2,p_3)$
are given in \ref{TableInt}

We direct the $z$--axis along the vector $\vQ$.
The transformation properites of the integrands in Eqs.(\ref{Glm})--(\ref{Fmu3})
under rotaions about the $z$--axis allow one to conclude that
\be\begin{array}{clclclcl}
&G^{l'm'}_{lm}(n)             &=&  \delta_{m',m}, &\qquad&
F^{l'm'}_{lm\mu_1}            &=&  \delta_{m',m+\mu},
\\
&G^{l'm'}_{lm\mu}(n)          &=&  \delta_{m',m+\mu}, &&
F^{l'm'}_{lm\mu_1\mu_2}       &=&  \delta_{m',m+\mu_1+\mu_2},
\\&                           & &         &&
F^{l'm'}_{lm\mu_1\mu_2\mu_3}  &=&  \delta_{m',m+\mu_1+\mu_2+\mu_3}.
\end{array}\ee

We express the components of the vector $\vec d=\vpp-\vp$ through
the spherical harmonics with the help of the relation
$p_\kappa=\sqrt{4\pi/3} \left|\vp\,\right| Y_{1\kappa}(\theta_p,\phi_p).$
Separating the depedence on the azimuthal angle
we write the spherical harmonics in terms of the
normalized associated Legendre polynomials
$
Y_{lm}\left(\theta,\phi\right)=
1/\sqrt{2\pi} \:
\exp(im\phi) \: \bar{P}^m_l(\cos\theta).
$

Being invariant under rotation of the coordinate system,
the functions $v(p_n)$ and $H(p_2,p_3)$ depend on
$\vp \cdot \vQ$, $\vpp \cdot \vQ$ and $\vp \cdot \vpp$
or equivalently on $p,p',Q,x,x'$ and $\cos(\phi-\phi')$,
where $x$ and $x'$ are the cosines of the polar angles
of the vectors $\vp$ and $\vpp$.

So, we arrive at the integrals of the type (we display here the most complicated one)
\be\label{FMPLM5}
\begin{array}{lll}\lefteqn{
{\dss\int\limits_{-1}^{+1} dx \int\limits_{-1}^{+1} dx'} \:
\bar{P}^{m'}_{l'}(x')  \bar{P}^m_l(x)
} &&\\&\times&
\bar{P}^{\mu_1}_1(x_1)  \bar{P}^{\mu_2}_1(x_2) \bar{P}^{\mu_3}_1(x_3)
R^M(p,p',Q;x,x')
\end{array}\ee
with
\be\label{RMExp5}
\begin{array}{lll}\lefteqn{
R^M(p,p',Q; x,x')={\dss\int\limits_0^{2\pi} d\phi \: \int\limits_0^{2\pi} d\phi'}
} &&\\&\times&
\exp[iM(\phi-\phi')]   f(p,p',Q;x,x',\cos(\phi-\phi')).
\end{array}\ee
In formulae (\ref{FMPLM5}) and (\ref{RMExp5}) we denote a linear combination
of $m',m,\mu_1,\mu_2,\mu_3$ by $M$,
the variables $x_1,x_2$ and $x_3$
should be identified with $x$ or $x'$.

Introducing the new integration variables
$t=\phi-\phi'$ and $t'=\phi+\phi'$ we get
\be\label{RMCos}
R^M(p,p',Q;x,x')=2\:
\int\limits^{2\pi}_{0} \: dt \: ( 2\pi -  t )
\cos(Mt) f(p,p',Q;x,x',\cos t).
\ee

The explicit expressions for the functions $F$ and $G$
in the form of Eq.(\ref{FMPLM5}) can be found in
\ref{TensorsFG}

It should be noted that the above procedure can be also used in the case of other models of MEC.
The next step depends on the specific form of the current operator.

For MEC given by (\ref{Jsgpion})--(\ref{vpi}) and the strong form factor chosen in the form
\be\label{FpiNN}
F_{\pi NN} \left(k^2\right) = \left(\Lpi-\mpi\right) {\left/ \right.}
                              \left(\Lpi+k^2\right),
\ee
the function $f(p,p',Q;x,x',\cos t)$ can be decomposed into a sum of pole terms
\be\label{PoleTerm}
\left\{\left[\Oh\vQ\pm\left(\vp-\vpp\right)\right]^2+\mu^2\right\}^{-n} =
(a+b\cos t)^{-n}, \qquad (n=1 \:\mbox{and}\: 2),
\ee
where $\mu=m_\pi$ or $\Lambda_\pi$.
In its turn, function (\ref{RMCos}) can be written (see Apendix E for details)
as a superposition of the integrals
\be\label{VabmnInt}                                                
V(a,b,m,n)=
\int\limits^{2\pi}_{0}  \: dt \: (2\pi-t)(a+b\cos t)^{-n} \cos(mt)
\ee
calculated for integer $n$ and $M$ in a closed form (\ref{VabmnClosedF}).
The two--dimensional integration over $x,x'$ in (\ref{FMPLM5}))
(and respectively, in Eqs. (\ref{GlmPV})--(\ref{Fmu3PW}))
is to be performed numerically.

Thus, we see that the analytical integration over the azimuthal angles
is feasible and one end up with 2-fold integrals of the type (\ref{FMPLM5})
instead of the 4-fold ones (\ref{M4Cyclic}).

For other models of MEC when the form of the matrix
elements of the current operator is not convenient to perform the
integration analytically one can utilize a procedure like that elaborated
in \cite{KSh87}. One can approximate
a function (which may be given on a mesh) by a  superposition of pole terms like
\be\label{PoleApproxim}
 F (k^2) \simeq \sum\limits_i \frac{C_i}{(k^2+m^2_i)^{n_i}}, \qquad (n_i=1,2,\ldots).
\ee
We have applied this procedure in \cite{KSh87} to
parameterize the solution of the Faddeev equations for the 3N bound
state. Separable nucleon--nucleon potentials have been used in
\cite{KSh87} to facilitate the computations.
These analytical integrations for $\phi $ and $\phi'$ reduce
the 4-fold integrals (\ref{M4Cyclic}) to 2-fold ones
of the type (\ref{FMPLM5}).

Another way to reduce the integrals (\ref{Glm})--(\ref{Fmu3})
is to write the integrand in terms of the spherical functions
and then apply the Clebsch--Gordan decomposition
\be\label{ClbGrdYD}
\begin{array}{lll}\lefteqn{
Y_{l_1 m_1}(\hat p) \; Y_{l_2 m_2}(\hat p) =
\sum_{LM} \; \left[
{{(2l_1+1)(2l_2+1) \over {4\pi (2L+1)}}}
\right]^{1/2}
} &&\\&\times&
C(l_1 l_2 L, 0 0 0) \; C( l_1 l_2 L , m_1, m_2, M)  \; Y_{L M}(\hat p).
\end{array}\ee
for products of spherical harmonics of the same argument.

As a result, one can get integrals in the form
\be\label{RLMCG}
\begin{array}{lll}\lefteqn{
R^{L'M'}_{LM}\left(p,p'; \vQ \right)
} &&\\&=&
{\dss\int \: d\hat p \: \int \: d\hat {p}'} \:
Y_{L'M'}^*({\hat p}') g\left(\vp,\vpp,\vQ\right) Y_{LM}(\hat p),
\end{array}\ee
where the function $g$ is a scalar.

One can see that
$ R^{L'M'}_{LM}\left(p,p'; \vQ \right) \sim \delta_{M'M} $
in the coordinate system where $\vQ\sim\vec e_x.$
Therefore, instead of Eq.(\ref{FMPLM5}) one can have
\be\label{FMPLM}
{\dss\int\limits_{-1}^{+1} dx \int\limits_{-1}^{+1} dx'} \:
P^M_L(x)  P^M_{L'}(x') R^M (p,p',Q;x,x').
\ee

Certainly, this way has much in common with the one described above
and in \ref{TensorsFG}
We have chosen the first of them for implementation in the
numerical calculations.

\section{Numerical Benchmarks}\label{NUMERI}
\setcounter{equation}0
\vskip -2ex

We have introduced three schemes to obtain the matrix elements 
$I_2$ in (\ref{EQI2}), e.g., the PWD from Appendix B, 
the 4FI based on (\ref{EQ4F}) and the 2FI via (\ref{FMPLM}). 
In this section we  
demonstrate that each scheme works well and the results agree perfectly. 
We choose as a test case the $\pi$-meson exchange two-body current (\ref{Jsgpion})--(\ref{Jpion}). The parameters we use are 
given in Table 1.
Further we choose Q=1 fm$^{-1}$ and put $F_1 ^V =1$ for the sake of simplicity. 
The PWD requires the double integrations in $x$ and $y$ according to (\ref{xyINT})
and (\ref{xyINT2}). Each integral is evaluated using 30 Gauss-Legendre points.
The methods 4FI and 2FI are evaluated by using for each integral variable 10 
Gauss-Legendre points. Also we drop the isospin part totally. Thus the 
quantity to be evaluated is for the spherical $\zeta$-component : 
\[
  \tilde I_2(p,p',Q; (l's') j' \mu ' , (l s) j \mu ) 
=  \sum _{m,m'} C (l s j , \mu -m , m ) C (l' s' j' , \mu ' -m' , m' )
\] 
\[
\times \sum _{k \kappa}
\int d \hat p d\hat {p'} Y^* _{l' \mu'-m' } (\hat {p'})
\left[ O^{k \kappa }  (\vec p_2 , \vec p_3 ) \right]_{\zeta}  Y_ {l \mu -m } (\hat p ) 
\left\langle s' m' \left| \left\{ 
\sigma (2) \otimes \sigma (3) \right\}_{k\kappa} \right| s m
\right\rangle
\]
\bl{rll}\label{EQI23}
\el
We list in Table 2  $\tilde I_2 $ for  typical two-body channel 
contributions and two sets of $p'$ and $p$ values. 
The three schemes  produce the same numbers within the 5 digits given. 
Note the change required in isospin due to  (\ref{ISOPART}).
Clearly one can not deduce any physics from that table, but 
it should be very useful for other groups for the purpose of checking 
their codes.
\begin{table}
\caption{ Parameters for the pionic current.
\label{table1}}
\begin{tabular}{lcccc}
$\hbar c  $  [MeV/fm]  &  $m_\pi$ [MeV] & $1/m_N$ [MeVfm$^2$] &
  $f^2 _{\pi NN} $
&  $\Lambda _\pi$ [MeV]   \\
\hline
197.33 &
139. & 41.467 & 0.081  & 1200. \\
\end{tabular}
\end{table}

\begin{table}
\caption{The benchmark quantity (\ref{EQI23}) for specific cases.
\label{table2}}
\begin{tabular}{lcccccc}
Final  state ($\mu'$)   & Initial state ($\mu$)  & $p'$ [fm$^{-1}$] 
 & $p$ [fm$^{-1}$]& $\tilde I_2^{seagull}$ [fm$^3$] & 
$\tilde I_2^{pionic}$[fm$^3$]  \\      
\hline
$^1$P$_1$(1) & $^1$S$_0$( 0) & 0.5 & 0.5 & -0.43878$\times $10$^{-1}$ &  0.12322$\times $10$^{-1}$ \\
$^3$S$_1$(1) & $^1$S$_0$( 0) & 0.5 & 0.5 & -0.69538$\times $10$^{-1}$ &  0.20792$\times $10$^{-1}$ \\
$^3$D$_1$(1) & $^1$S$_0$( 0) & 0.5 & 0.5 &  0.53929$\times $10$^{-2}$ &  0.53001$\times $10$^{-2}$ \\
$^1$P$_1$(1) & $^3$P$_0$( 0) & 0.5 & 0.5 & -0.13068$\times $10$^{-5}$ &  0.76770$\times $10$^{-2}$ \\
$^3$S$_1$(1) & $^3$P$_0$( 0) & 0.5 & 0.5 & -0.69148$\times $10$^{-1}$ &  0.17510$\times $10$^{-1}$ \\
$^3$D$_1$(1) & $^3$P$_0$( 0) & 0.5 & 0.5 & -0.21208$\times $10$^{-2}$ & -0.15741$\times $10$^{-1}$ \\
$^1$S$_0$(0) & $^1$P$_1$(-1) & 0.5 & 0.5 & -0.43878$\times $10$^{-1}$ &  0.12322$\times $10$^{-1}$ \\
$^3$P$_0$(0) & $^1$P$_1$(-1) & 0.5 & 0.5 & -0.13068$\times $10$^{-5}$ &  0.76770$\times $10$^{-2}$ \\
$^3$P$_1$(0) & $^1$P$_1$(-1) & 0.5 & 0.5 &  0.75258$\times $10$^{-2}$ &  0.17771$\times $10$^{-1}$ \\
$^3$P$_1$(1) & $^1$P$_1$( 0) & 0.5 & 0.5 &  0.25989$\times $10$^{-1}$ & -0.18852$\times $10$^{-2}$ \\
$^1$S$_0$(0) & $^3$S$_1$(-1) & 0.5 & 0.5 & -0.69538$\times $10$^{-1}$ &  0.20792$\times $10$^{-1}$ \\
$^3$P$_0$(0) & $^3$S$_1$(-1) & 0.5 & 0.5 & -0.69148$\times $10$^{-1}$ &  0.17510$\times $10$^{-1}$\\
$^3$P$_1$(0) & $^3$S$_1$(-1) & 0.5 & 0.5 &  0.63478$\times $10$^{-1}$ & -0.77629$\times $10$^{-2}$ \\
$^3$P$_1$(1) & $^3$S$_1$( 0) & 0.5 & 0.5 &  0.54138$\times $10$^{-1}$ & -0.20495$\times $10$^{-1}$ \\
$^1$P$_1$(0) & $^3$P$_1$(-1) & 0.5 & 0.5 & -0.25989$\times $10$^{-1}$ &  0.18852$\times $10$^{-2}$ \\
$^1$P$_1$(1) & $^3$P$_1$( 0) & 0.5 & 0.5 & -0.75258$\times $10$^{-2}$ & -0.17771$\times $10$^{-1}$ \\
$^3$S$_1$(0) & $^3$P$_1$(-1) & 0.5 & 0.5 & -0.54138$\times $10$^{-1}$ &  0.20495$\times $10$^{-1}$ \\ 
$^3$S$_1$(1) & $^3$P$_1$( 0) & 0.5 & 0.5 & -0.63478$\times $10$^{-1}$ &  0.77629$\times $10$^{-2}$  \\
$^3$D$_1$(0) & $^3$P$_1$(-1) & 0.5 & 0.5 &  0.21553$\times $10$^{-2}$ &  0.10227$\times $10$^{-1}$ \\
$^3$D$_1$(1) & $^3$P$_1$( 0) & 0.5 & 0.5 & -0.30655$\times $10$^{-2}$ &  0.11589$\times $10$^{-1}$  \\
$^1$S$_0$(0) & $^3$D$_1$(-1) & 0.5 & 0.5 &  0.53929$\times $10$^{-2}$ &  0.53001$\times $10$^{-2}$ \\
$^3$P$_0$(0) & $^3$D$_1$(-1) & 0.5 & 0.5 & -0.21208$\times $10$^{-2}$ & -0.15741$\times $10$^{-1}$ \\
$^3$P$_1$(0) & $^3$D$_1$(-1) & 0.5 & 0.5 &  0.30655$\times $10$^{-2}$ & -0.11589$\times $10$^{-1}$ \\
$^3$P$_1$(1) & $^3$D$_1$( 0) & 0.5 & 0.5 & -0.21553$\times $10$^{-2}$ & -0.10227$\times $10$^{-1}$ \\
%
%
$^1$P$_1$(1) & $^1$S$_0$( 0) & 1.0 & 0.5 & -0.57838$\times $10$^{-1}$ &  0.30361$\times $10$^{-1}$ \\
$^3$S$_1$(1) & $^1$S$_0$( 0) & 1.0 & 0.5 & -0.34083$\times $10$^{-1}$ &  0.22421$\times $10$^{-1}$ \\
$^3$D$_1$(1) & $^1$S$_0$( 0) & 1.0 & 0.5 &  0.12077$\times $10$^{-1}$ &  0.11928$\times $10$^{-2}$ \\
$^1$P$_1$(1) & $^3$P$_0$( 0) & 1.0 & 0.5 & -0.13462$\times $10$^{-5}$ &  0.64118$\times $10$^{-2}$ \\
$^3$S$_1$(1) & $^3$P$_0$( 0) & 1.0 & 0.5 & -0.33735$\times $10$^{-1}$ &  0.11544$\times $10$^{-1}$ \\
$^3$D$_1$(1) & $^3$P$_0$( 0) & 1.0 & 0.5 & -0.21350$\times $10$^{-2}$ & -0.25089$\times $10$^{-1}$ \\
$^1$S$_0$(0) & $^1$P$_1$(-1) & 1.0 & 0.5 & -0.19222$\times $10$^{-1}$ &  0.16863$\times $10$^{-1}$ \\
$^3$P$_0$(0) & $^1$P$_1$(-1) & 1.0 & 0.5 & -0.46181$\times $10$^{-5}$ &  0.64118$\times $10$^{-2}$ \\
$^3$P$_1$(0) & $^1$P$_1$(-1) & 1.0 & 0.5 &  0.46799$\times $10$^{-2}$ &  0.91221$\times $10$^{-2}$ \\
$^3$P$_1$(1) & $^1$P$_1$( 0) & 1.0 & 0.5 &  0.18041$\times $10$^{-1}$ & -0.31948$\times $10$^{-2}$ \\
$^1$S$_0$(0) & $^3$S$_1$(-1) & 1.0 & 0.5 & -0.34083$\times $10$^{-1}$ &  0.22421$\times $10$^{-1}$ \\
$^3$P$_0$(0) & $^3$S$_1$(-1) & 1.0 & 0.5 & -0.95407$\times $10$^{-1}$ &  0.27836$\times $10$^{-1}$\\
$^3$P$_1$(0) & $^3$S$_1$(-1) & 1.0 & 0.5 &  0.84423$\times $10$^{-1}$ & -0.19717$\times $10$^{-1}$ \\
$^3$P$_1$(1) & $^3$S$_1$( 0) & 1.0 & 0.5 &  0.76826$\times $10$^{-1}$ & -0.32340$\times $10$^{-1}$ \\
$^1$P$_1$(0) & $^3$P$_1$(-1) & 1.0 & 0.5 & -0.18037$\times $10$^{-1}$ &  0.31948$\times $10$^{-2}$ \\
$^1$P$_1$(1) & $^3$P$_1$( 0) & 1.0 & 0.5 & -0.46799$\times $10$^{-2}$ & -0.91221$\times $10$^{-2}$ \\
$^3$S$_1$(0) & $^3$P$_1$(-1) & 1.0 & 0.5 & -0.28138$\times $10$^{-1}$ &  0.13312$\times $10$^{-1}$ \\ 
$^3$S$_1$(1) & $^3$P$_1$( 0) & 1.0 & 0.5 & -0.28423$\times $10$^{-1}$ &  0.11098$\times $10$^{-1}$  \\
\end{tabular}
\end{table}

To illustrate the behaviour of the currents we show in Figs. 2-6 examples 
for $\tilde I_2 $ for the seagull and pionic parts and their sum.
We choose again typical channel combinations and 
threshold behaviors  $ \sim {p'} ^{l'} p^l $ can be seen. 
Please note the different shapes of the currents and the different magnitudes of the pionic and seagull parts.

\begin{figure}[t]
(a)
\psfig{file=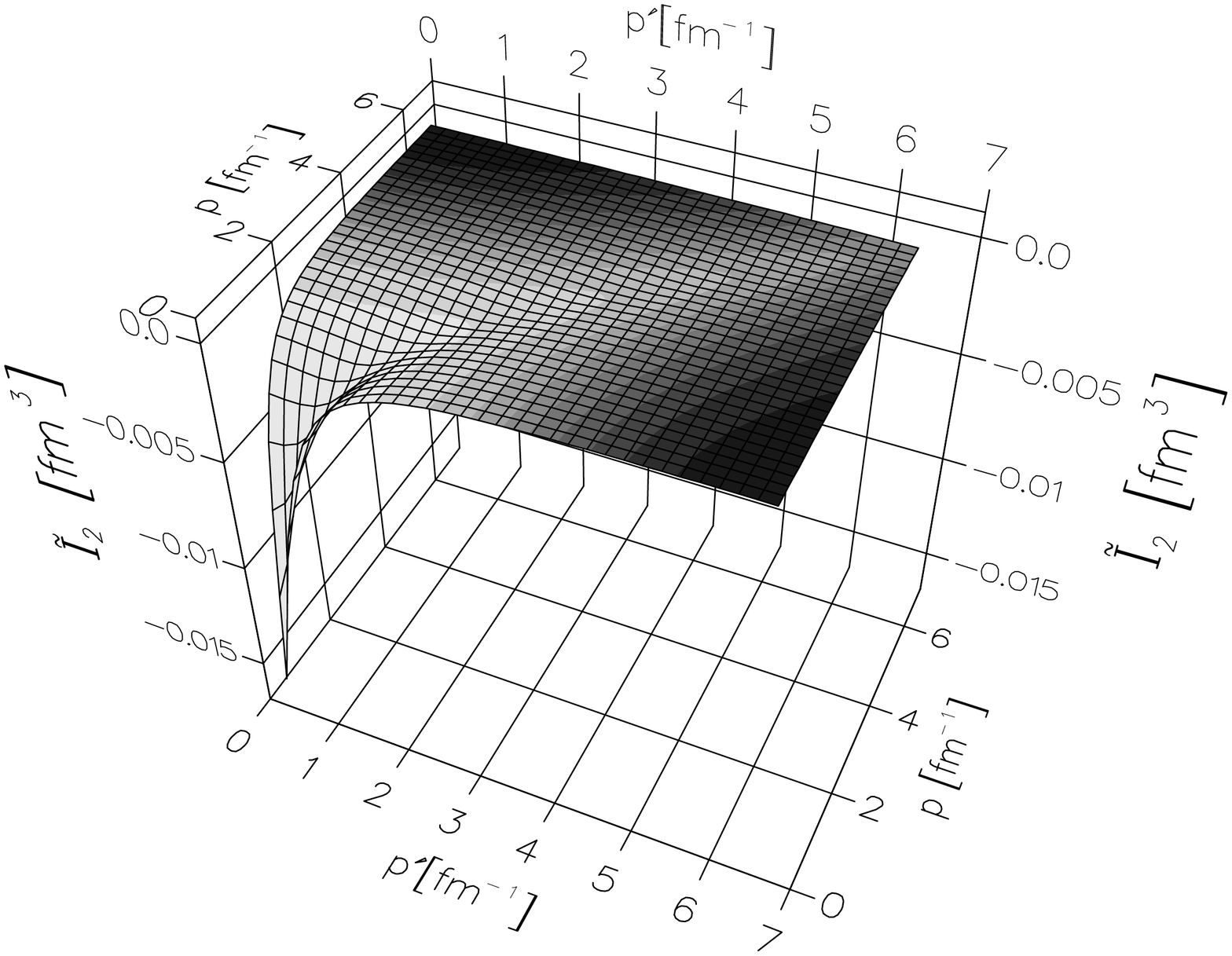,angle=-90,scale=0.35}

(b)

\psfig{file=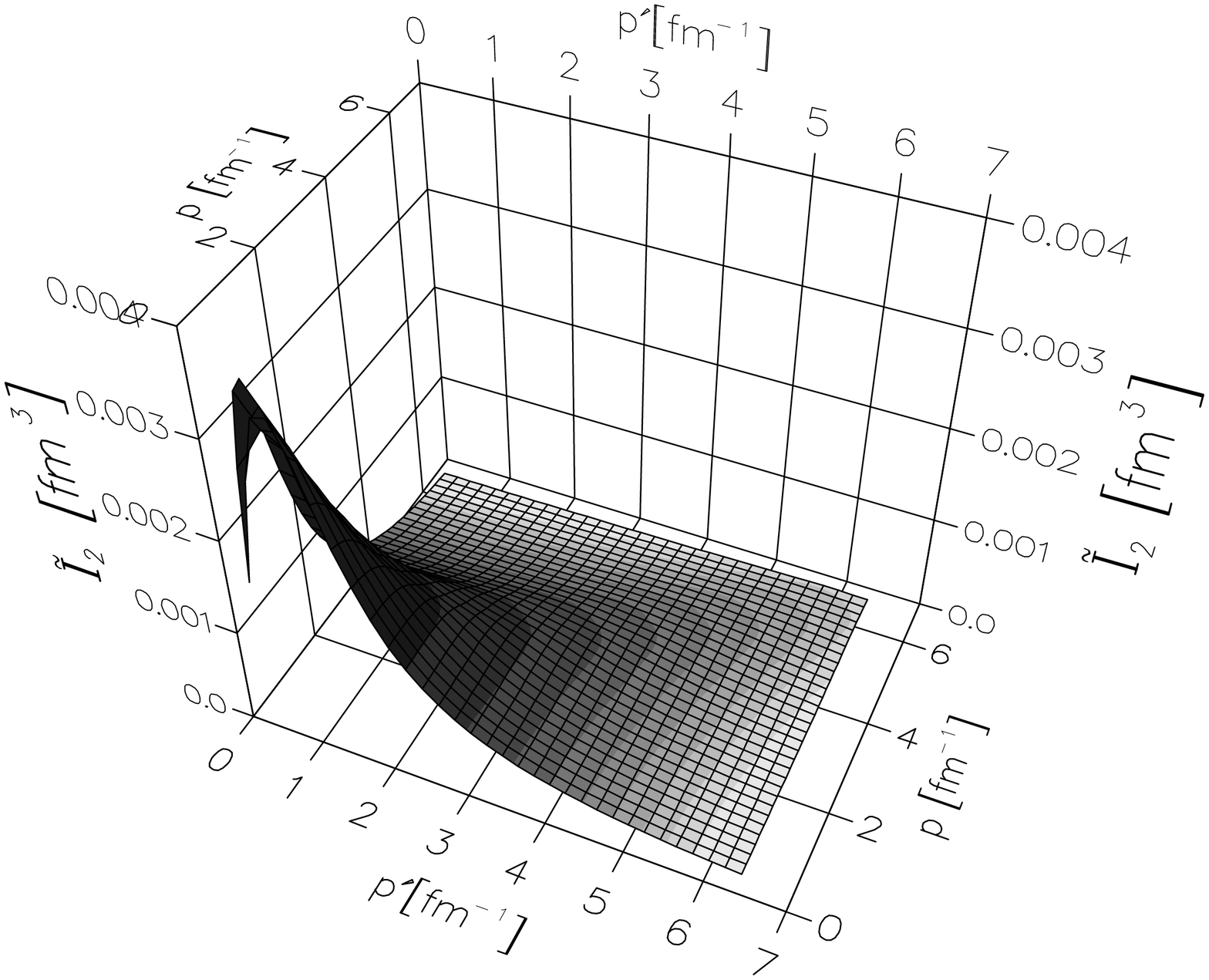,angle=-90,scale=0.35}

(c)
\psfig{file=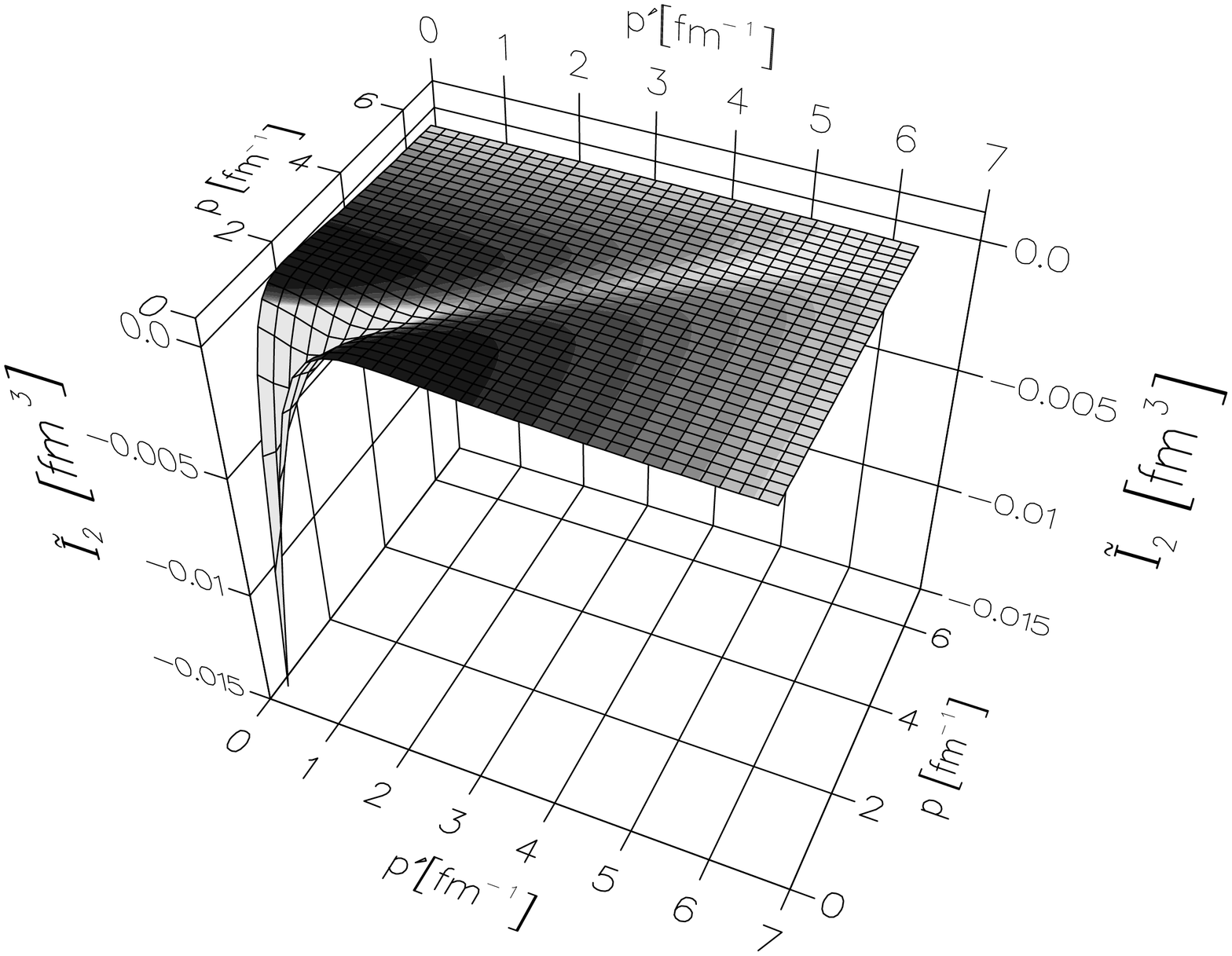,angle=-90,scale=0.35}
\caption{The expression $\tilde I_2$ from Eq. (\ref{EQI23}), (a) the seagull term,
(b) the pionic term and (c) the sum of both.
The  initial and final states are $^1$S$_0$($\mu$=0) and $^3$S$_1$($\mu '$ =1),
respectively.}
\label{fig1}
\end{figure}

\begin{figure}[t]
(a)
\psfig{file=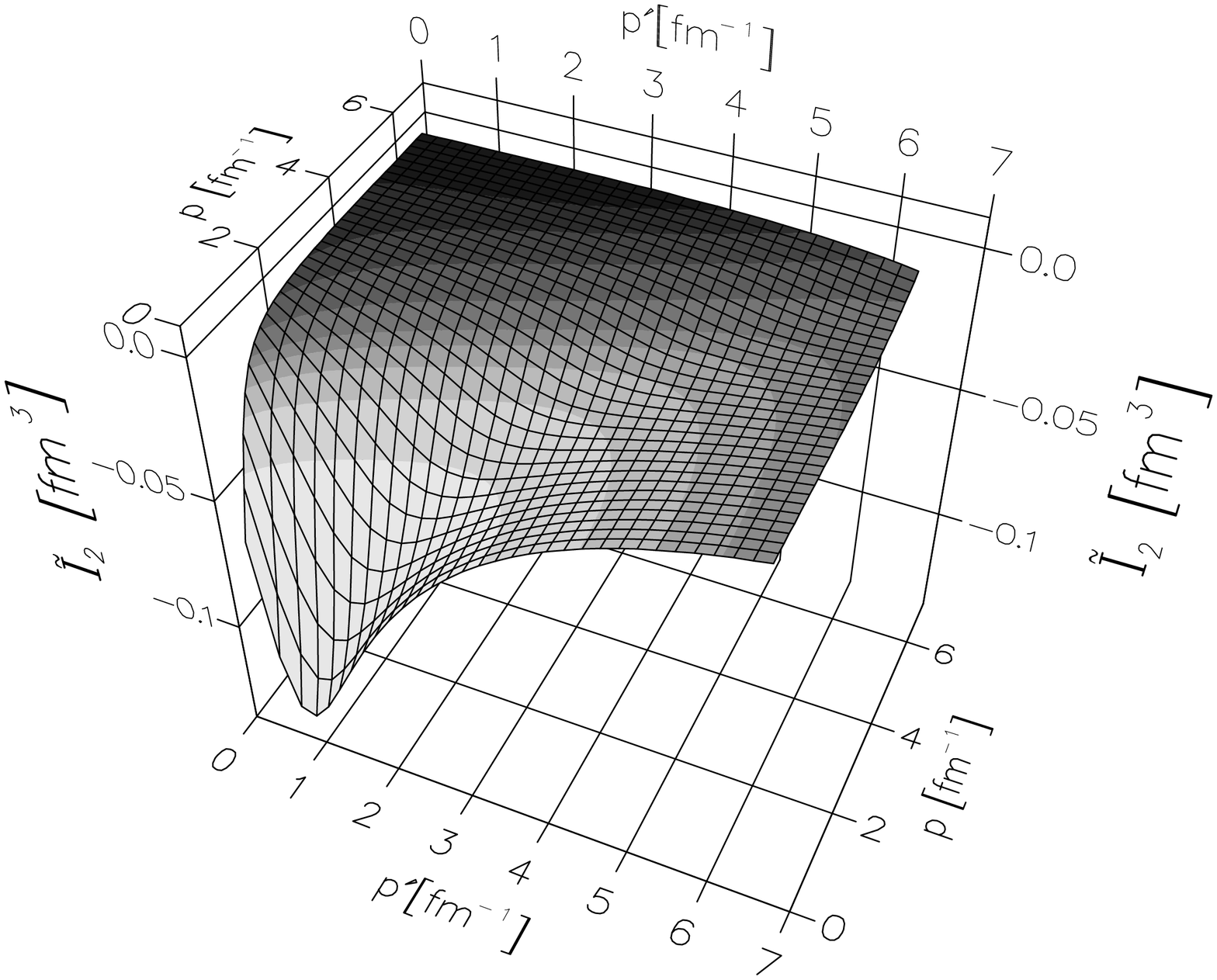,angle=-90,scale=0.35}

(b)
\psfig{file=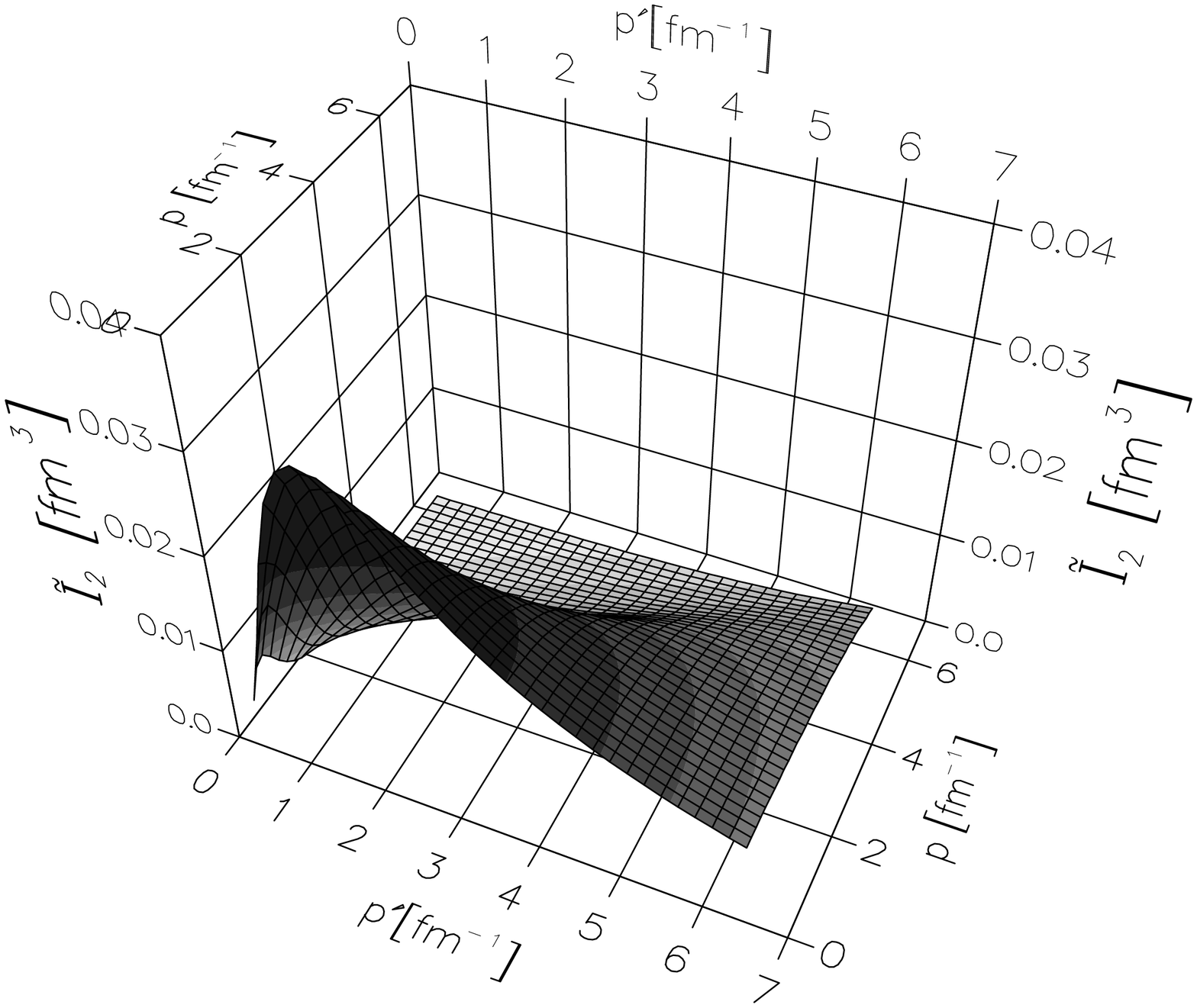,angle=-90,scale=0.35}

(c)
\psfig{file=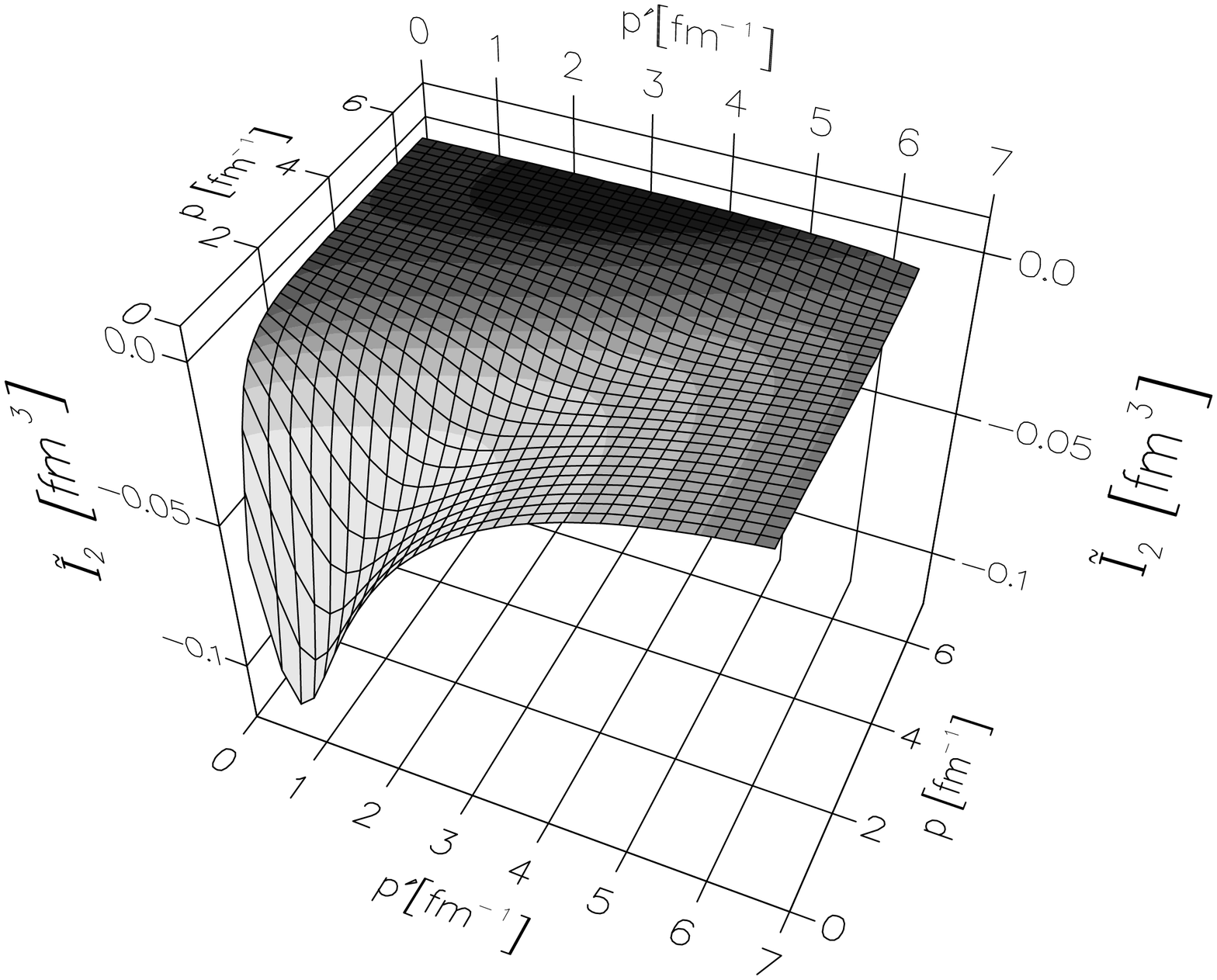,angle=-90,scale=0.35}
\caption{The same as in Fig. \ref{fig1} for
$^3$P$_0$($\mu =0$)  and  $^3$S$_1$($\mu'=1$).
}
\label{fig2}
\end{figure}

\begin{figure}[t]
(a)
\psfig{file=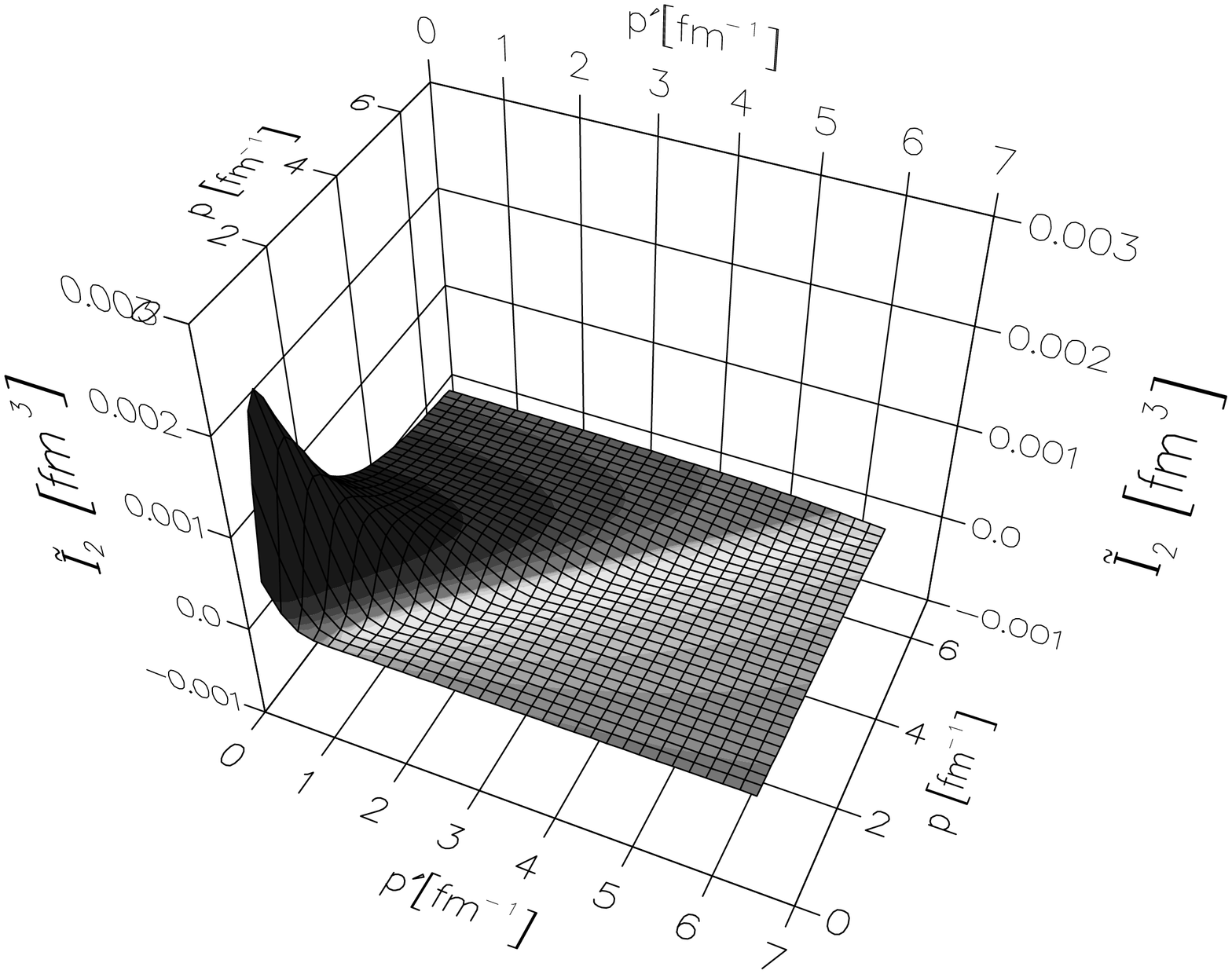,angle=-90,scale=0.35}

(b)
\psfig{file=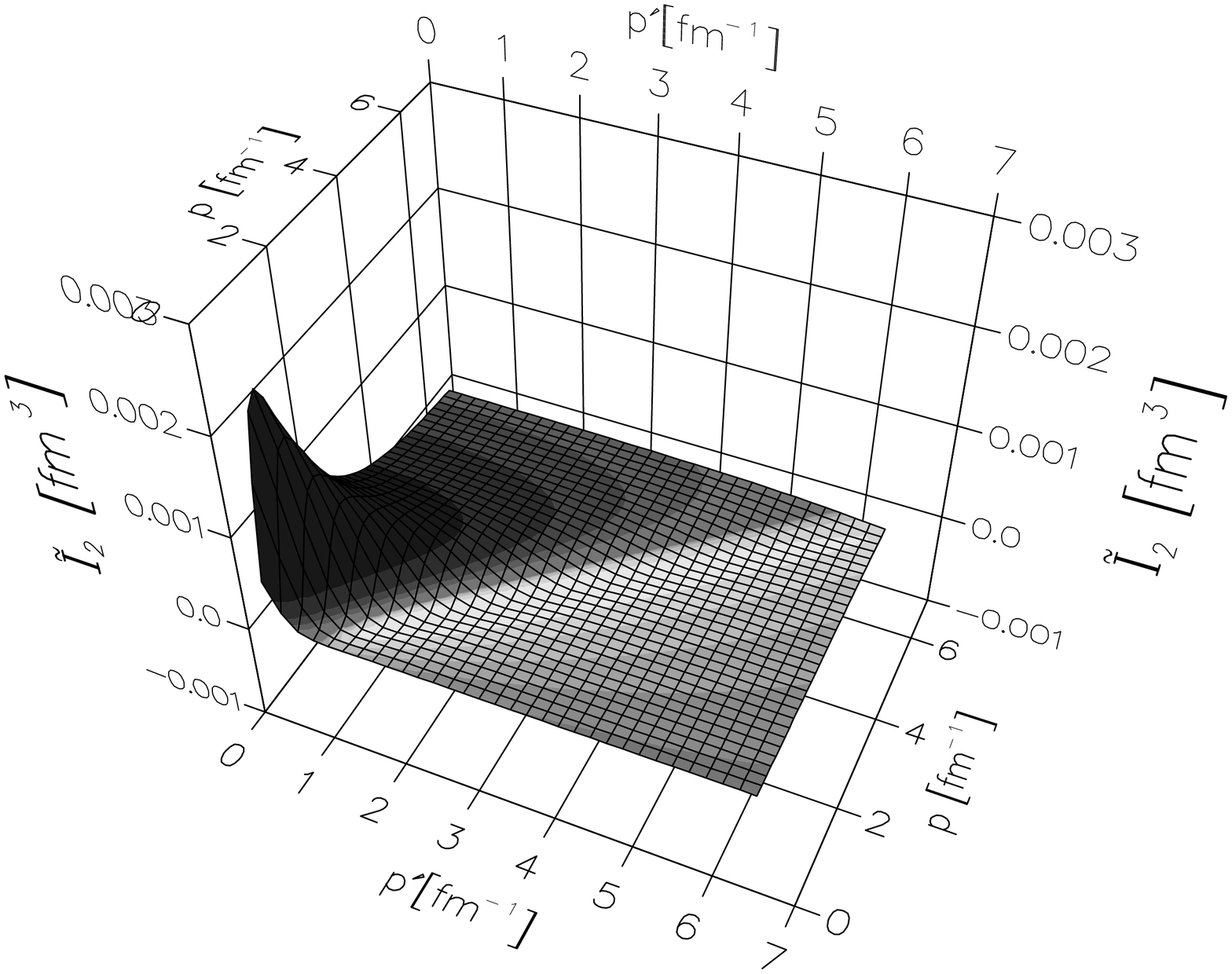,angle=-90,scale=0.35}

(c)
\psfig{file=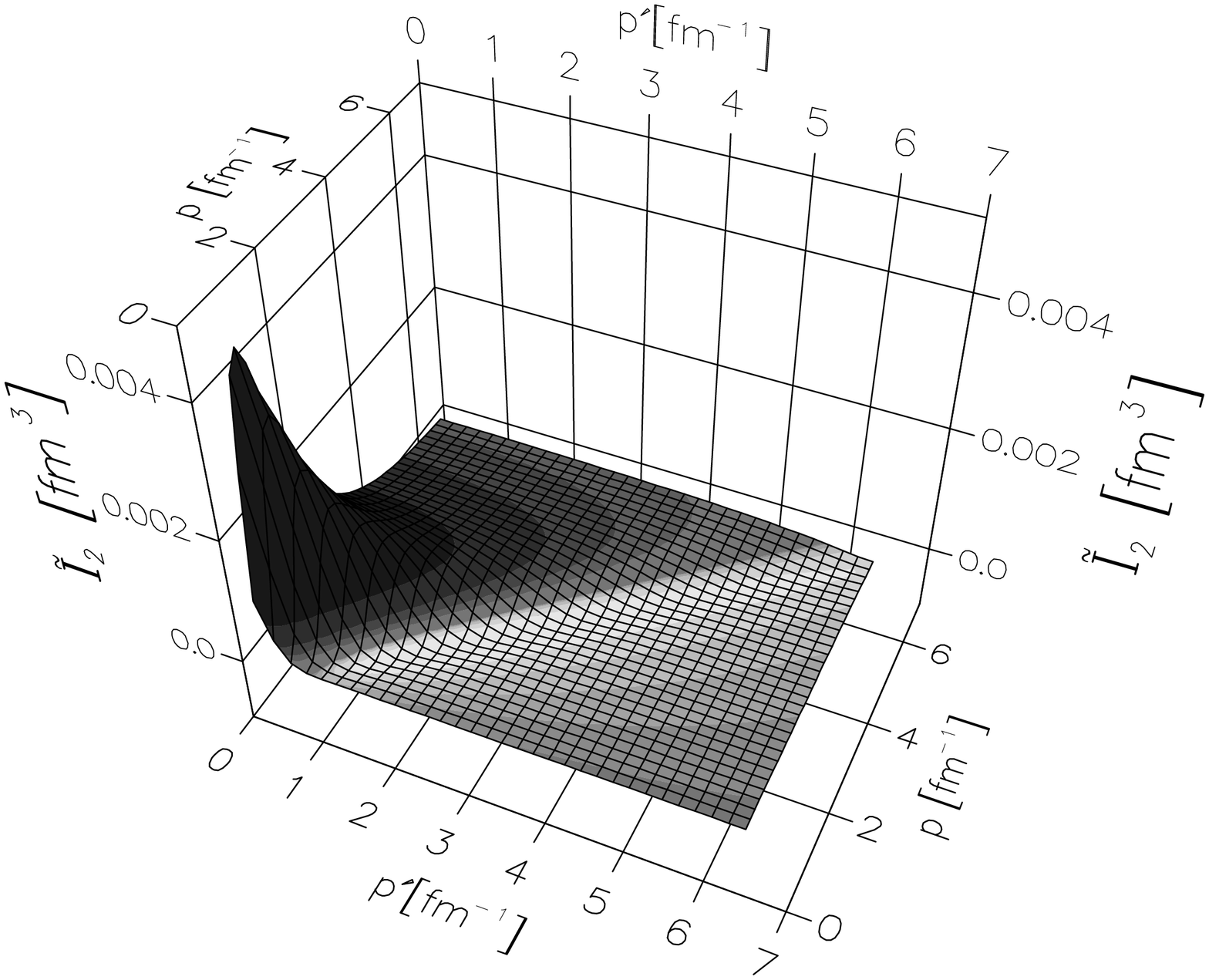,angle=-90,scale=0.35}
\caption{The same as in Fig. \ref{fig1} for
$^1$S$_0$($\mu=0$) and $^3$D$_1$($\mu'$=1).
}
\label{fig3}
\end{figure}

\begin{figure}[t]
(a)
\psfig{file=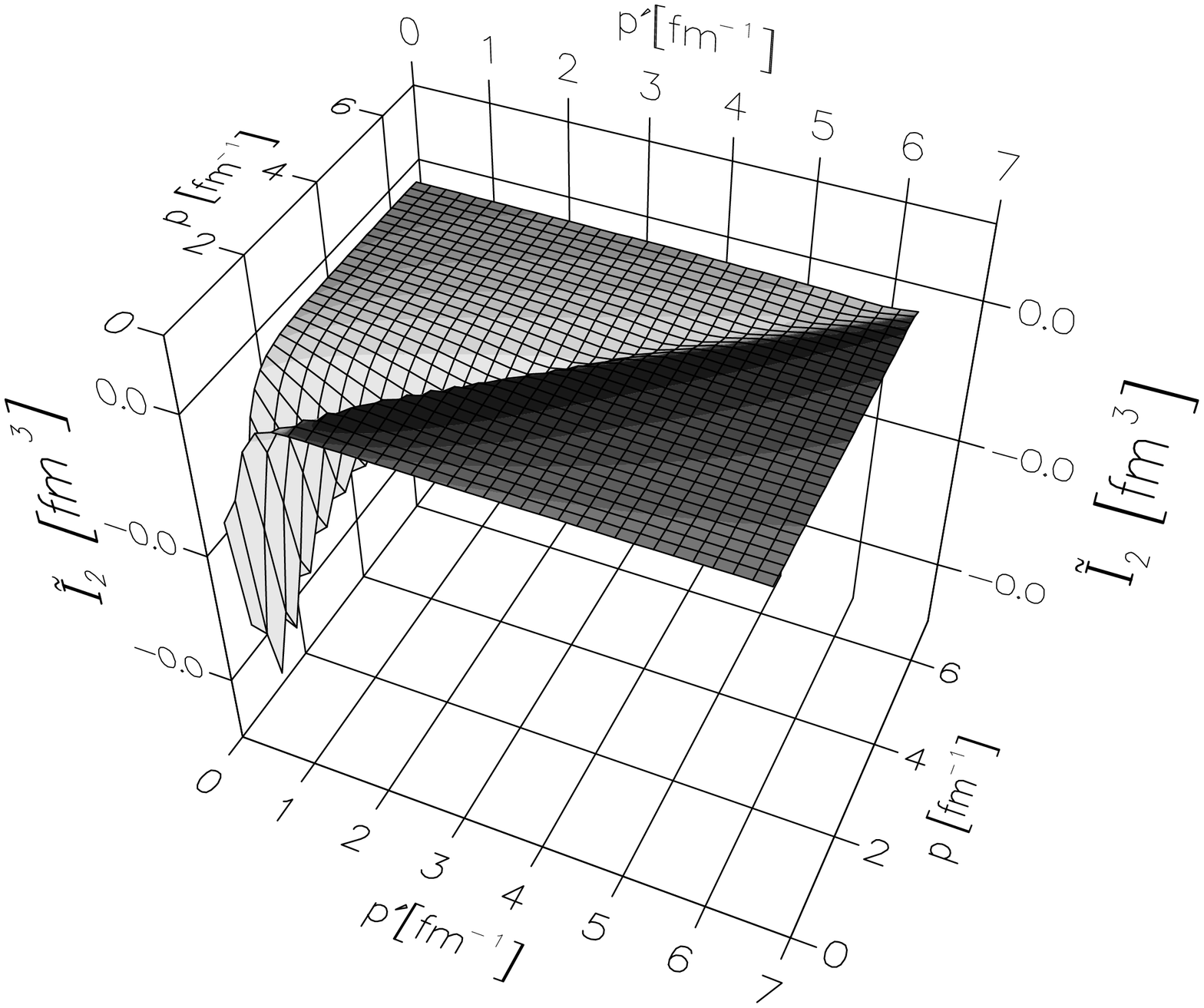,angle=-90,scale=0.35}

(b)
\psfig{file=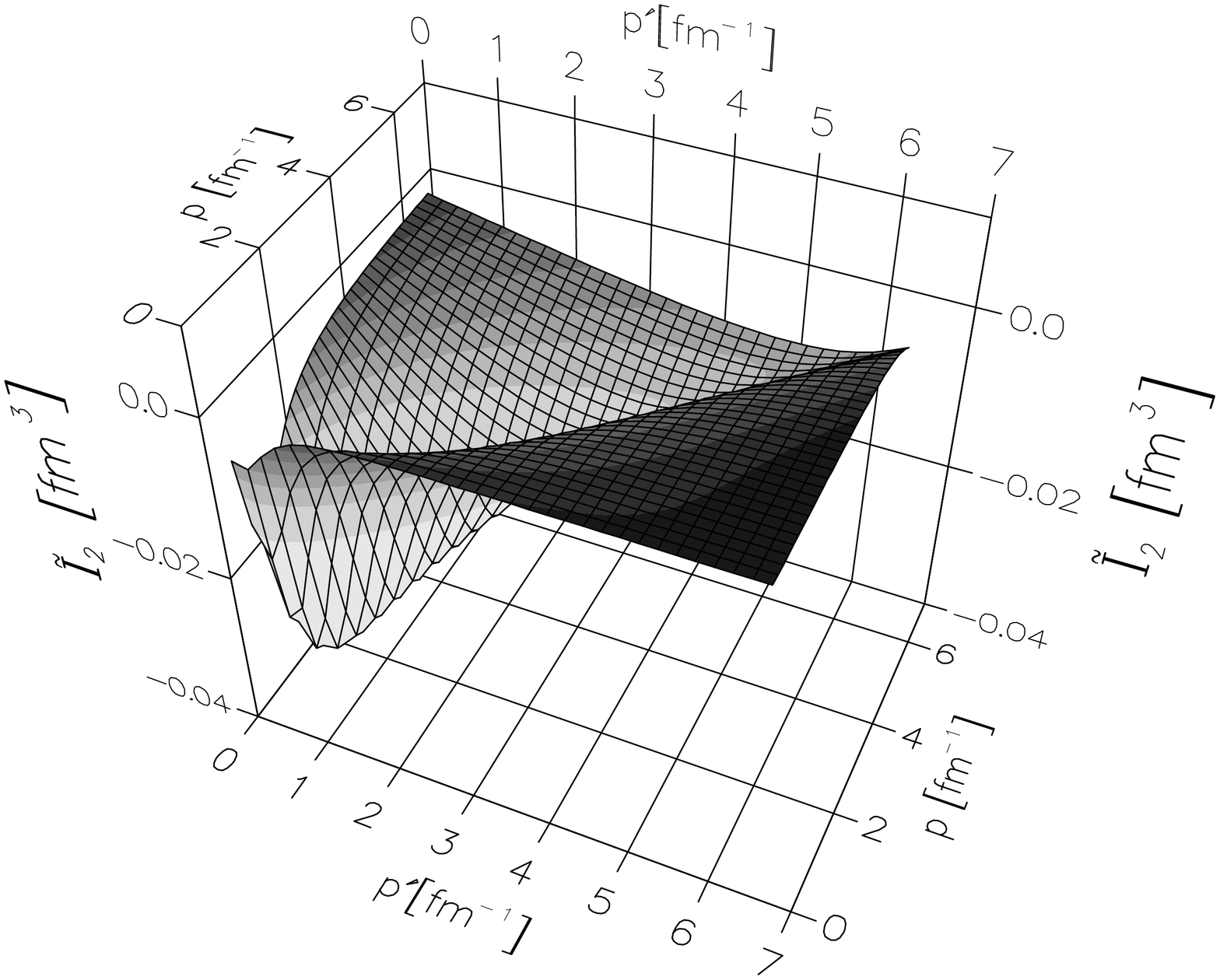,angle=-90,scale=0.35}

(c)
\psfig{file=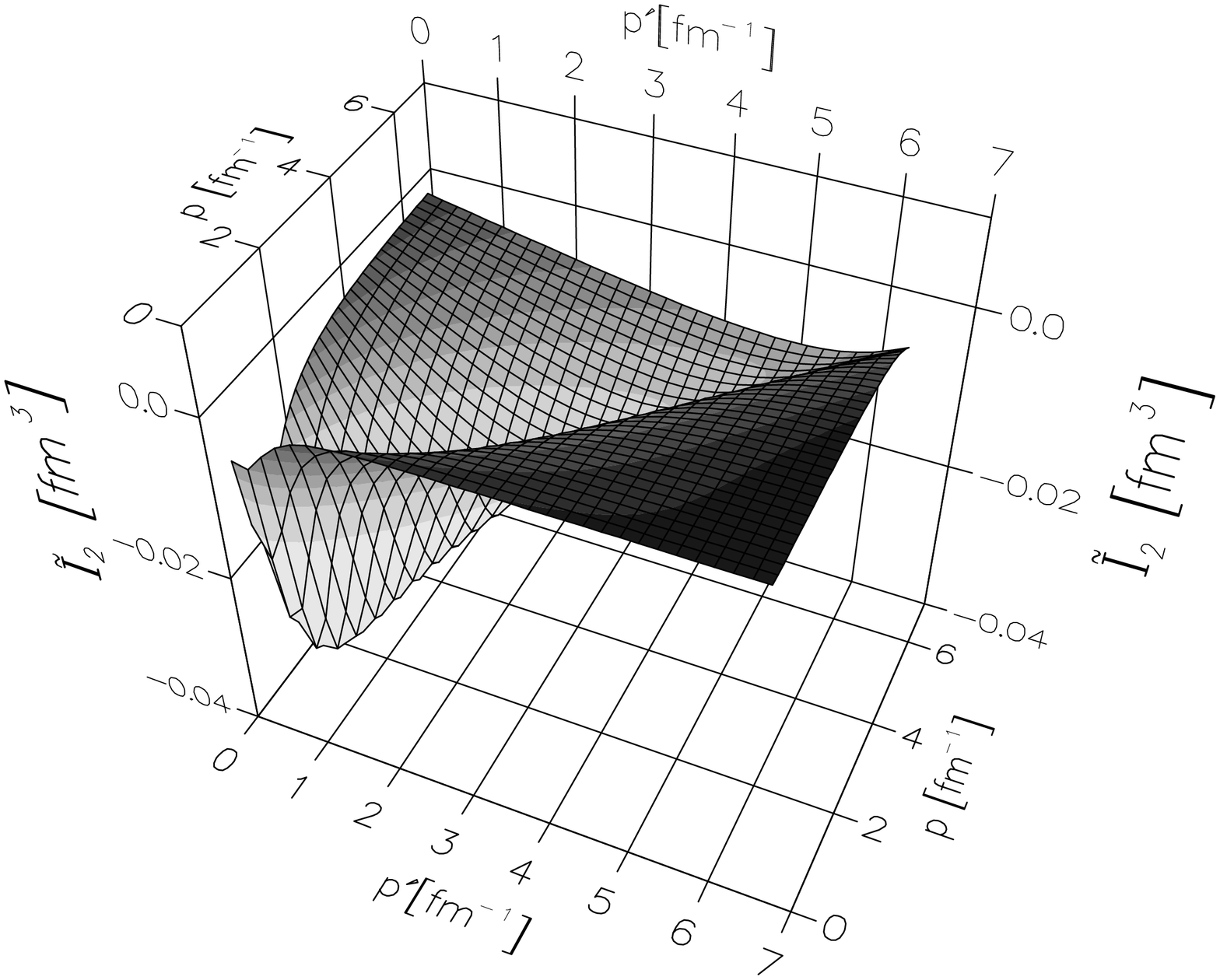,angle=-90,scale=0.35}
\caption{
The same as in Fig. \ref{fig1} for
$^3$P$_0$($\mu$=0) and $^3$S$_1$($\mu'$=1).
}
\label{fig4}
\end{figure}

\begin{figure}[t]
\psfig{file=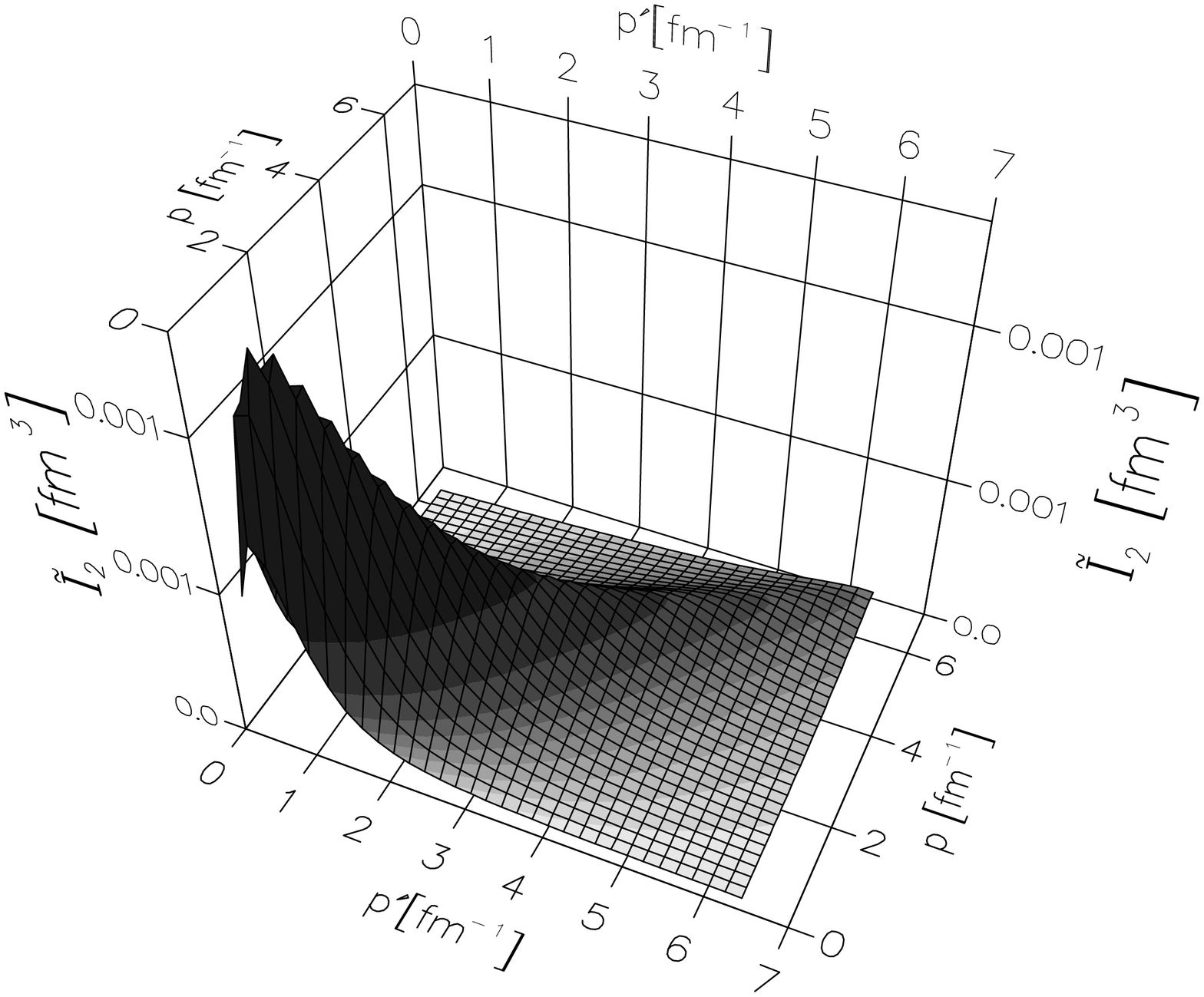,angle=-90,scale=0.35}
\caption{
The expression $\tilde I_2$ from (\ref{EQI23}), only the pionic part
contributes. 
The initial and final states are $^3$P$_0$($\mu$=0) and $^1$P$_1$($\mu'$=1).
}
\label{fig5}
\end{figure}

\section{Summary and Discussion}\label{Conclusions}
\setcounter{equation}0
\vskip -2ex

      
In this paper we develop three approaches (PWD, 2FI and 4FI)
for calculating the matrix elements of the two--nucleon EM currents.
We use model \cite{Riska11},\cite{Mathiot173} of MEC and
take into account the contributions originating form $\pi$-- and $\rho$--exchanges.
The considered MEC satisfy the continuity equation with the one--pion--one--rho--meson
exchange potential.

To get the matrix elements of interest we apply the current operator to
the 3N bound state and project this vector on 3N basis states $\left|pq\alpha\right>$.
The latter are specified by the magnitudes of the Jacobi momenta $p,q$
and the discrete quantum numbers $\alpha$ in $jJ$--coupling.
This separation in Eqs.(\ref{EQI2})--(\ref{EQI3}) of both the operation
and the projection can in general work on N-nucleon systems.

The matrix elements in question are contained as building blocks in
the amplitudes of the \He breakup by photons and electrons.
We choose this particular form of the matrix elements
since it is convenient for incorporation of the corresponding results
into the calculations of the reaction amplitudes
with full inclusion of the final state interaction.

We express the matrix elements in terms of the multiple integrals.
Resorting to the technique of the partial wave decompositions we reduce the matrix elements
and arrive at the expressions convenient for the numerical calculations.

For example, the matrix elements of MEC in $\left|pq\alpha\right>$ basis have the form of
a superposition of four--dimensional angular integrals.
We elaborate two different ways (PWD and 2FI) to diminish the dimensionality of integration
and end up with two--fold integrals. One of them, 2FI, is outlined in Section 3.
Expressions coming from PWD are given explicitly in Appendices B and C
for $\pi$-- and $\rho$--MEC, respectively.

Performing the integration over the azimuthal angles in the closed forms (see Section 3),
we face tedious analytical calculations. To carry out the second of these integrations
one needs to exploit the explicit form of the MEC. This means, in particular,
that each model of MEC requires a special treatment within the 2FI procedure.

Aiming to provide a flexibility of the approach, we work out a third procedure (4FI) to
compute the discussed four-fold integrals with numerical methods.

We would like to note, that the feasibility of the direct numerical calculation of the
reaction amplitudes represented in a form of the multidimensional integrals has been
demonstrated in \cite{KShS}--\cite{Spin96},\cite{KSh87}.
The six--dimensional integrals evaluated there have much in common with the ones
we deal in this work.
In fact, in the both cases an essential element in the integrands is
the mixed tensor  ${\vec O}^{k\kappa} \left(\vec p_2, \vec p_3 \right)$
arising from decomposition of the two--body current.
The Korobov method \cite{Korobov63} of numerical integration in many dimensions
have been applied in Refs.\cite{KShS}--\cite{Spin96} and \cite{KSh87}.
This method proved to be an effective tool for computation of the six-fold integrals.

The detailed analysis show that the results obtained with
independent methods (PWD, 2FI and 4FI) are in the perfect agreement.
In Section 4 we present benchmark results for the $\pi$--MEC.
The momentum dependence of the matrix elements MEC in $\left|pq\alpha\right>$ basis is
studied as well.

The potentialities of the approaches elaborated are not restricted by the
consideration of the simple two--body currents generated by the $\pi$-- and $\rho$--exchange.
Other, more complicated, models of the MEC consistent with the realistic NN potentials
can be treated without any substantial modifications of the methods.

So, we calculate the matrix elements of the  two--body currents in a form
providing the compatibility with approach \cite{IshikawaA107}--\cite{Golak97}.
The results of the present article are to be used in forthcoming investigations of the
quasi--elastic electron scattering off \He and proton--deuteron radiative capture
with inclusion both the MEC and the final state interaction.
As it is discussed above,
these elements of the theory have to be treated simultaneously
for ensuring the gauge independence of the reaction amplitudes and, respectively,
for obtaining unambiguous predictions for the observables.

\bigskip
{\bf Acknowledgment }
One of the authors (V. V. K.)
would like to thank A.V. Shebeko for the helpful discussions.
This work was supported financially (V. V. K.) by the
German Ministry for Education, Science and Technology (BMBF No. RUB19960),
the Polish Committee for Scientific Research (J. G.) ( grant No. 2P03B03914)
and the Deutsche Forschungsgemeinschaft (H. K.).
The numerical calculations have been performed on the CRAY T90
and CRAY T3E of the John von Neumann Institute for Computing, J\"ulich, Germany.


\appendix
\setcounter{section}0
\renewcommand{\thesection}{Appendix \Alph{section}.}
\renewcommand{\theequation}{\Alph{section}.\arabic{equation}}

\setcounter{equation}0
\section{The partial wave representation of Eq. (\ref{EQI3})}
\label{APPI3}

Eq. (\ref{EQI3}) reads 
\be 
I _3( q',q,Q ; (\lambda' {1\over2}) I' M'-\mu ' ,( \lambda {1 \over
2}) I M- \mu ) \equiv
 \int d\hat {q'} {\cal Y } _{I', M'-\mu '} ^* ( \hat {q'} )
{{\delta ( q -\vert \vec {q'} + {1 \over 3} \vec Q \vert ) } \over { q } ^2 }
{\cal Y } _{I, M-\mu}  ( \widehat {\vec {q'} + {1 \over 3} \vec Q} ).
\ee
We use the relations 
\be
{{ \delta ( q - \left| \vec {q'} + {1 \over 3}\vec Q \right| ) } 
\over{ q ^{\lambda + 2 }} } = 
\sum _k 2\pi (-)^k \sqrt{ \hat k } h_{k\lambda}(q',q,Q) {\cal Y } _{kk}^
{00} (\hat {q'} , \hat Q)  
\ee
with 
\be
h_{k \lambda}(q',q,Q) \equiv 
\int _{-1} ^ 1 dx P_k ( x) {{ \delta ( q - \sqrt{ {q'}^2 + Q^2/9+ 2q'Q x /3} )}
\over { q ^{ \lambda + 2 }}} 
\ee
and 
\be
Y_{lm}(\widehat{ \vec a + \vec b }) =\sum _{ l_1 +l_2 = l}
{ {a^{l_1} b^{l_2}} \over { \vert \vec a + \vec b \vert ^ l } } 
\sqrt{ { 4 \pi (2 l + 1 )! } \over 
{ (2 l_1 +1 ) ! (2 l_2 + 1 )! } }
{\cal Y }_{l_1 l_2}^{l m}(\hat a , \hat b )
\ee

Then assuming that the direction of $\vec Q $ is parallel to the z-
axis
we get employing standard steps
\bl{rl}
& I_3 ( q',q,Q ; (\lambda' {1\over2}) I' M'-\mu ' ,( \lambda {1 \over
2}) I M- \mu ) =  \delta_{m'-\mu ' , m-\mu }  { 1 \over 2 } 
\sqrt{ \hat \lambda \hat I } (-) ^{ \lambda + I' + { 1 \over 2}} 
\sqrt{ (2\lambda +1 )! }
\\ & \times 
\sum_{ \lambda _1 + \lambda _2 = \lambda } { q ' } ^{\lambda_1 } ( { 1 \over
3 } Q ) ^{ \lambda _2 } { 1 \over { \sqrt{ (2 \lambda _1)! (2\lambda_2)! }}}
\sum_k \hat k (-)^k h_k (q',q,Q) C ( \lambda k \lambda ' ,0 0 0 ) 
\\ & \times 
\sum_f \sqrt{\hat f } C(\lambda _2 k f, 0 0 0 ) 
\left\{ \begin{array}{ccc}
\lambda '  & f  & \lambda  \\ \lambda _2 & \lambda _1 & k 
\end{array} \right\}
\left\{ \begin{array}{ccc}
 f &  \lambda  & \lambda ' \\ { 1 \over 2 } & I '  & I
\end{array} \right\}
C( f,I,I',0 , m-\mu  ). 
\el

\setcounter{equation}0
\section{The partial wave representation of Eq. (\ref{EQI2})}
\label{APPI2}
Here we show a spherical component $\zeta$ of $\vec I_2 $ and drop the 
isospin part. This corresponds to Eq. (\ref{EQI23}). 
 The photon momentum $ \hat Q $ is assumed to point into the z-direction.
 We evaluated the seagull part  (\ref{Jseag})
and the pionic one  (\ref{Jpion}).

\[
\tilde I_2 ^{seagull} ( p',p,Q;(l's')j'\mu', (ls)j\mu ) 
\]
\[ = - \ 6 \sqrt{3} \pi 
     \sqrt{ \hat{l} \hat{s} \hat{j} \hat{l'} \hat{s'} \hat{j'} }
(-1)^{l'+s'} \, \delta_{{\mu}',{\mu}+\zeta} \ 
\]
\[ \times
\sum\limits_{\alpha_1} (-1)^{\alpha_1} \hat{{\alpha_1}} 
  \ \left \{ { \begin{array}{ccc}
                       1  &1 &  {\alpha_1}  \\
                   \frac12   & \frac12 & s  \\
                    \frac12  & \frac12 & s'
                   \end{array}} \right \}
\]
\[ \times
\sum\limits_{\alpha_2} (-1)^{\alpha_2} C( 1 j \alpha_2 ; \zeta \mu {\mu}') 
\]
\[ \times
\sum\limits_{\alpha_3} (-1)^{\alpha_3}  \hat{{\alpha_3}} \ 
                 \left \{ { \begin{array}{ccc}
                       1  &  {\alpha_1} & 1 \\
                       l  &    s        & j \\
                     {\alpha_3} & s' &  {\alpha_2} 
                   \end{array}} \right \}
\]
\[ \times
\sum\limits_{h_1 + h_2 = 1} (-1)^{h_2} ( \frac12 Q )^{h_1} 
\]
\[ \times
\sum\limits_{r} \hat{r} [ 1 - (-1)^{\alpha_1 + h_2 + r} ]
\]
\[ \times
\sum\limits_{f_1} \sqrt{\hat{f_1}} C( r h_1 f_1 ; 0 0 0 ) C(f_1 j' \alpha_2 ; 0 {\mu}' {\mu}') 
                 \left \{ { \begin{array}{ccc}
                       f_1  &    l'       & \alpha_3 \\
                       s' & {\alpha_2} & j'
                   \end{array}} \right \}
\]
\[ \times
\sum\limits_{f_2} \sqrt{\hat{f_2}} C( r h_2 f_2 ; 0 0 0 ) \sqrt{ ( 2 f_2 + 1 ) ! } 
                 \left \{ { \begin{array}{ccc}
                       f_1  &    f_2   & 1 \\
                       l & {\alpha_3} & l'
                   \end{array}} \right \} \
                 \left \{ { \begin{array}{ccc}
                       f_2  &    f_1   & 1 \\
                       h_1 & h_2 & r
                   \end{array}} \right \}
\]
\[ \times
\sum\limits_{u_1 + u_2 = f_2} (p')^{\, u_1} (p)^{u_2} \frac{1}{ \sqrt{ ( 2 u_1 + 1 ) !  ( 2 u_2 ) ! }}
\]
\begin{equation}
\times
\sum\limits_{z} \sqrt{\hat{z}} C( u_2 l z ; 0 0 0 ) C( l' z u_1 ; 0 0 0 ) 
                 \left \{ { \begin{array}{ccc}
                       u_1  &    u_2   & f_2 \\
                       l & l' & z
                   \end{array}} \right \} \ G^{h_2 f_2}_{z r} {\rm ,} 
\label{jpiI}
\end{equation}
where 
\begin{equation}\label{xyINT}
 G^{h_2 f_2}_{z r} = \int\limits_{-1}^1 d y P_z (y) \Delta^{ h_2 - f_2 } \, 
                     \int\limits_{-1}^1 d x P_r (x) 
          v_\pi ( \frac14 Q^2 + \Delta^2 + Q \Delta x ) \ \ 
  {\rm and} \ \ \Delta = \sqrt{ p^2 + {p'}^{\, 2} - 2 p p' y} {\rm .}
\end{equation}

\[
\tilde I_2 ^{pionic} (p', p Q; (l's')j' \mu ', (ls)j \mu ) 
\]
\[ = 36 \pi 
     \sqrt{ \hat{l} \hat{s} \hat{j} \hat{l'} \hat{s'} \hat{j'} }
(-1)^{l+s'} \, \delta_{{\mu}',{\mu}+\zeta} \ 
\]
\[   \times
\sum\limits_{\alpha_1} (-1)^{\alpha_1} \hat{{\alpha_1}} 
  \ \left \{ { \begin{array}{ccc}
                       1  &1 &  {\alpha_1}  \\
                   \frac12   & \frac12 & s  \\
                    \frac12  & \frac12 & s'
                   \end{array}} \right \}
\]
\[  \times
\sum\limits_{e} \hat{e}
\]
\[  \times
\sum\limits_{\alpha_2} (-1)^{\alpha_2} C( 1 j \alpha_2 ; \zeta \mu {\mu}') 
\]
\[  \times
\sum\limits_{\alpha_3} (-1)^{\alpha_3}  \hat{{\alpha_3}} \ 
                 \left \{ { \begin{array}{ccc}
                       1  &  {\alpha_1} & 1 \\
                       l  &    s        & j \\
                     {\alpha_3} & s' &  {\alpha_2} 
                   \end{array}} \right \}
\]
\[  \times
\sum\limits_{h_1 + h_2 = 1} (-1)^{h_1} ( \frac12 Q )^{h_2} 
\]
\[  \times
\sum\limits_{w_1 + w_2 = 1} ( \frac12 Q )^{w_2} 
\]
\[  \times
\sum\limits_{d_1} \sqrt{\hat{d_1}} C( h_1 w_1 d_1 ; 0 0 0 ) 
\]
\[  \times
 \sum\limits_{d_2} \sqrt{\hat{d_2}} C( h_2 w_2 d_2 ; 0 0 0 ) 
  \ \left \{ { \begin{array}{ccc}
                       h_1  & h_2 &  1  \\
                       w_1  & w_2 &  1  \\
                       d_1  & d_2 &  \alpha_1 
                   \end{array}} \right \}
\]
\[  \times
\sum\limits_{c} \sqrt{\hat{c}} C( d_1 1 c ; 0 0 0 ) 
  \ \left \{ { \begin{array}{ccc}
                       d_2  & d_1 &  \alpha_1 \\
                       1  & e &  c 
                   \end{array}} \right \}
\]
\[  \times
\sum\limits_{f_1} \sqrt{\hat{f_1}} 
                 \left \{ { \begin{array}{ccc}
                       f_1  &    l'       & \alpha_3 \\
                       s' & {\alpha_2} & j'
                   \end{array}} \right \}
       (-1)^{f_1} \, C(f_1 j' \alpha_2 ; 0 {\mu}' {\mu}') 
\]
\[  \times
\sum\limits_{f_2} \sqrt{\hat{f_2}} 
                 \left \{ { \begin{array}{ccc}
                       f_1  &    f_2   & e \\
                       l & {\alpha_3} & l'
                   \end{array}} \right \} \
\sqrt{ ( 2 f_2 + 1 ) ! } 
\]
\[ \times
\sum\limits_{r} \hat{r} 
C( r d_2 f_1 ; 0 0 0 )
C( r c f_2  ; 0 0 0 )
                 \left \{ { \begin{array}{ccc}
                       d_2  &    c   & e \\
                       f_2 & f_1 & r
                   \end{array}} \right \} \
\]
\[ \times
\sum\limits_{u_1 + u_2 = f_2} (p')^{\, u_1} (p)^{u_2} 
   \frac{1}{ \sqrt{ ( 2 u_1 + 1 ) !  ( 2 u_2 ) ! }}
\]
\begin{equation}
\times
\sum\limits_{z} \sqrt{\hat{z}} C( u_2 l z ; 0 0 0 ) C( l' z u_1 ; 0 0 0 ) 
                 \left \{ { \begin{array}{ccc}
                       u_1  &    u_2   & f_2 \\
                       l & l' & z
                   \end{array}} \right \} \ H^{h_1 w_1 f_2}_{z r} {\rm ,} 
\label{jpiII}
\end{equation}
where 
\begin{equation}\label{xyINT2}
 H^{h_1 w_1 f_2}_{z r} = \int\limits_{-1}^1 d y P_z (y) 
                  \Delta^{ 1 + h_1 + w_1 - f_2 } \, 
                     \int\limits_{-1}^1 d x P_r (x) 
  \frac { v_\pi ( \frac14 Q^2 + \Delta^2 - Q \Delta x )
       -  v_\pi ( \frac14 Q^2 + \Delta^2 + Q \Delta x )}
        { 2 Q \Delta x } {\rm .}
\end{equation}
We used the abbreviation $\hat a \equiv 2 a + 1 $.


\setcounter{equation}0
\section{The $\rho$ Meson Exchange Currents }
\label{APPRHO}

The $\rho$-meson exchange current
are given for instance in \cite{Riska11} and 
\cite{Mathiot173}.
The current consists of four parts.

\[
{\vec j}_{\rho} ( {\vec p}_2 , {\vec p}_3 ) \ = \ 
{\vec j}_{\rho , I} ( {\vec p}_2 , {\vec p}_3 ) \ +
{\vec j}_{\rho , II} ( {\vec p}_2 , {\vec p}_3 ) \ +
{\vec j}_{\rho , III} ( {\vec p}_2 , {\vec p}_3 ) \ +
{\vec j}_{\rho , IV} ( {\vec p}_2 , {\vec p}_3 ) \ 
\]

\[
{\vec j}_{\rho, I} ( {\vec p}_2 , {\vec p}_3 ) \ \equiv \  
  i [ {\vec \tau}(2) \times {\vec \tau}(3) ]_z \ F_1^V ( Q^2 ) \
\frac{ {\vec p}_2 - {\vec p}_3 }{ p_2^2 - p_3^2 } \
\left( v_\rho^s (p_3) - v_\rho^s (p_2) \right)
\]

\[
{\vec j}_{\rho, II} ( {\vec p}_2 , {\vec p}_3 ) \ \equiv \  
 - \  i [ {\vec \tau}(2) \times {\vec \tau}(3) ]_z \ F_1^V ( Q^2 ) \
\]
\[ \times
\left( v_\rho (p_3) \, {\vec \sigma}(2) \times ( {\vec \sigma}(3) \times {\vec p}_3 ) 
 \ - \ v_\rho (p_2) \, {\vec \sigma}(3) \times ( {\vec \sigma}(2) \times {\vec p}_2 )  \right)
\]

\[
{\vec j}_{\rho, III} ( {\vec p}_2 , {\vec p}_3 ) \ \equiv \  
 - \  i [ {\vec \tau}(2) \times {\vec \tau}(3) ]_z \ F_1^V ( Q^2 ) \
\frac{ v_\rho (p_3) - v_\rho (p_2) }{ p_2^2 - p_3^2 } \
\left( ( {\vec \sigma}(2) \times {\vec p}_2 ) \cdot ( {\vec \sigma}(3) \times {\vec p}_3 ) \right)
 ( {\vec p}_2 - {\vec p}_3 )
\]
\[
{\vec j}_{\rho, IV} ( {\vec p}_2 , {\vec p}_3 ) \ \equiv \  
 - \  i [ {\vec \tau}(2) \times {\vec \tau}(3) ]_z \ F_1^V ( Q^2 ) \
\frac{ v_\rho (p_3) - v_\rho (p_2) }{ p_2^2 - p_3^2 } \
\]
\[ \times 
\left( {\vec \sigma}(3) \cdot ( {\vec p}_2 \times {\vec p}_3 ) ( {\vec \sigma}(2) \times {\vec p}_2 )\ + \ 
 {\vec \sigma}(2) \cdot ( {\vec p}_2 \times {\vec p}_3 ) ( {\vec \sigma}(3) \times {\vec p}_3 ) \right) {\rm .}
\]

First we present their PWD  for the spherical component $\zeta$ :
\[
\tilde I_2 ^{I \rho} ( p',p, Q  ; (l's')j' \mu ', (ls) j \mu ) 
\]

\[ = 2 \pi 
     \sqrt{ \hat{l} \hat{j} \hat{l'} }
(-1)^{l+j'} \, \delta_{s',s} \ \delta_{{\mu}',{\mu}+\zeta} \ 
\]
\[ \times
\sum\limits_{\alpha_2}  \sqrt{\hat{{\alpha_2}}} 
       (-1)^{\alpha_2} C( 1 j \alpha_2 ; \zeta \mu {\mu}') 
\]
\[  \times
\sum\limits_{\alpha_3} (-1)^{\alpha_3}  \hat{{\alpha_3}} \ 
                 \left \{ { \begin{array}{ccc}
                       1  &  l & {\alpha_3} \\
                       s  &   {\alpha_2} & j
                   \end{array}} \right \}
\]
\[ \times
\sum\limits_{r} \hat{r} 
                 \left \{ { \begin{array}{ccc}
                       r  &  {\alpha_3} & l' \\
                       s' & j' &  {\alpha_2}
                   \end{array}} \right \} \
C( r  {\alpha_2} j' ; 0 {\mu}' {\mu}' )
\]
\[ \times
\sum\limits_{\alpha_4} \sqrt{\hat{\alpha_4}} \, (-1)^{\alpha_4} \,
     \sqrt{ ( 2 \alpha_4 + 1 ) ! } C( r 1 \alpha_4 ; 0 0 0 ) \,
                 \left \{ { \begin{array}{ccc}
                       r  &  1 & {\alpha_4} \\
                       l & l' &  {\alpha_3}
                   \end{array}} \right \} \
\]
\[ \times
\sum\limits_{u_1 + u_2 = \alpha_4} (p')^{\, u_1} (p)^{u_2} (-1)^{u_2}
   \frac{1}{ \sqrt{ ( 2 u_1 + 1 ) !  ( 2 u_2 ) ! }}
\]
\begin{equation}
\times
\sum\limits_{z} \sqrt{\hat{z}} (-1)^{z} 
       C( u_2 l z ; 0 0 0 ) C( l' z u_1 ; 0 0 0 ) 
                 \left \{ { \begin{array}{ccc}
                       u_1  &    u_2   & \alpha_4 \\
                       l & l' & z
                   \end{array}} \right \} \ H^{\alpha_4}_{z r} {\rm ,} 
\label{jrhoI}
\end{equation}
where 
\begin{equation}
 H^{\alpha_4}_{z r} = \int\limits_{-1}^1 d y P_z (y) 
                  \Delta^{ 1 - \alpha_4 } \, 
                     \int\limits_{-1}^1 d x P_r (x) 
  \frac {v_\rho^s ( \frac14 Q^2 + \Delta^2 - Q \Delta x )
       - v_\rho^s ( \frac14 Q^2 + \Delta^2 + Q \Delta x )}
        { 2 Q \Delta x } {\rm .}
\end{equation}

\[
\tilde I_2 ^{II \rho} ( p',p , Q; (l's')j' \mu ', (ls) j \mu ) 
\]

\[ = 36 \pi \sqrt{3} \,
     \sqrt{ \hat{l} \hat{s} \hat{j} \hat{l'} \hat{s'} \hat{j'} }
(-1)^{s'} \, \delta_{{\mu}',{\mu}+\zeta } \
\]
\[  \times
\sum\limits_{\alpha_1} (-1)^{\alpha_1} \hat{{\alpha_1}} 
  \ \left \{ { \begin{array}{ccc}
                       1  &1 &  {\alpha_1}  \\
                       1 & 1 & 1
                   \end{array}} \right \}
  \ \left \{ { \begin{array}{ccc}
                       1  &1 &  {\alpha_1}  \\
                   \frac12   & \frac12 & s  \\
                    \frac12  & \frac12 & s'
                   \end{array}} \right \}
\]
\[  \times
\sum\limits_{u_1 + u_2 = 1} (-1)^{u_1} ( \frac12 Q )^{u_2} 
\]
\[  \times 
\sum\limits_{r} (-1)^{r} \hat{r} \, [ 1 - (-1)^{\alpha_1 + u_1 + r} ]
\]
\[  \times
\sum\limits_{f_1} \sqrt{\hat{f_1}}  (-1)^{f_1} \, \sqrt{ ( 2 f_1 + 1 ) ! } \,
C( r u_1 f_1 ; 0 0 0 )
\]
\[  \times
\sum\limits_{f_2} \sqrt{\hat{f_2}}  (-1)^{f_2} \,
                 \left \{ { \begin{array}{ccc}
                       f_2  &    f_1   & 1 \\
                       u_1 & u_2 & r
                   \end{array}} \right \} \
C( r u_2 f_2 ; 0 0 0 )
\]
\[  \times
\sum\limits_{\alpha_2} (-1)^{\alpha_2} C( 1 j \alpha_2 ; \zeta \mu {\mu}') \,
C( f_2 j' \alpha_2 ; 0 {\mu}' {\mu}') 
\]
\[  \times
\sum\limits_{\alpha_3} (-1)^{\alpha_3}  \hat{{\alpha_3}} \ 
                 \left \{ { \begin{array}{ccc}
                       1  &  {\alpha_1} & 1 \\
                       l  &    s        & j \\
                     {\alpha_3} & s' &  {\alpha_2} 
                   \end{array}} \right \} \
                 \left \{ { \begin{array}{ccc}
                       f_2  &  f_1 & 1 \\
                       l  &   \alpha_3    & l' 
                   \end{array}} \right \}
                 \left \{ { \begin{array}{ccc}
                       f_2  &  l' & \alpha_3 \\
                       s' &   \alpha_2    & j' 
                   \end{array}} \right \}
\]
\[  \times
\sum\limits_{w_1 + w_2 = f_1} (p')^{\, w_1} (p)^{w_2} (-1)^{w_2}
   \frac{1}{ \sqrt{ ( 2 w_1 + 1 ) !  ( 2 w_2 ) ! }}
\]
\begin{equation}
\times
\sum\limits_{z} \sqrt{\hat{z}} (-1)^z \,
           C( w_2 l z ; 0 0 0 ) C( l' z w_1 ; 0 0 0 ) 
                 \left \{ { \begin{array}{ccc}
                       w_1  &    w_2   & f_1 \\
                       l & l' & z
                   \end{array}} \right \} \ H^{u_1 f_1}_{z r} {\rm ,} 
\label{jrhoII}
\end{equation}
where 
\begin{equation}
 H^{u_1 f_1}_{z r} = - \int\limits_{-1}^1 d y P_z (y) 
                  \Delta^{ u_1 - f_1 } \, 
                     \int\limits_{-1}^1 d x P_r (x) 
  v_\rho ( \frac14 Q^2 + \Delta^2 - Q \Delta x ) {\rm .}
\end{equation}

\[
\tilde I_2 ^{III \rho} ( p',p, Q; (l's')j' \mu ', (ls) j \mu ) 
\]

\[ = -216 \pi 
     \sqrt{ \hat{l} \hat{s} \hat{j} \hat{l'} \hat{s'} \hat{j'} }
(-1)^{l+s'} \, \delta_{{\mu}',{\mu}+\zeta } \ =
\]
\[  \times
\sum\limits_{\alpha_1} \hat{{\alpha_1}} 
  \ \left \{ { \begin{array}{ccc}
                       1  &1 &  1  \\
                       1  &1 &  {\alpha_1} 
                   \end{array}} \right \}
  \ \left \{ { \begin{array}{ccc}
                       1  &1 &  {\alpha_1}  \\
                   \frac12   & \frac12 & s  \\
                    \frac12  & \frac12 & s'
                   \end{array}} \right \}
\]
\[  \times
\sum\limits_{e} \hat{e}
\]
\[  \times
\sum\limits_{\alpha_2} (-1)^{\alpha_2} C( 1 j \alpha_2 ; \zeta \mu {\mu}') 
\]
\[  \times
\sum\limits_{\alpha_3} (-1)^{\alpha_3}  \hat{{\alpha_3}} \ 
                 \left \{ { \begin{array}{ccc}
                       1  &  {\alpha_1} & 1 \\
                       l  &    s        & j \\
                     {\alpha_3} & s' &  {\alpha_2} 
                   \end{array}} \right \}
\]
\[  \times
\sum\limits_{h_1 + h_2 = 1} (-1)^{h_1} ( \frac12 Q )^{h_2} 
\]
\[  \times
\sum\limits_{w_1 + w_2 = 1} ( \frac12 Q )^{w_2} 
\]
\[  \times
\sum\limits_{d_1} \sqrt{\hat{d_1}} C( h_1 w_1 d_1 ; 0 0 0 ) 
\]
\[  \times
\sum\limits_{d_2} \sqrt{\hat{d_2}} C( h_2 w_2 d_2 ; 0 0 0 ) 
  \ \left \{ { \begin{array}{ccc}
                       h_1  & h_2 &  1  \\
                       w_1  & w_2 &  1  \\
                       d_1  & d_2 &  \alpha_1 
                   \end{array}} \right \}
\]
\[  \times
\sum\limits_{c} \sqrt{\hat{c}} C( d_1 1 c ; 0 0 0 ) 
  \ \left \{ { \begin{array}{ccc}
                       d_2  & d_1 &  \alpha_1 \\
                       1  & e &  c 
                   \end{array}} \right \}
\]
\[  \times
\sum\limits_{f_1} \sqrt{\hat{f_1}} 
                 \left \{ { \begin{array}{ccc}
                       f_1  &    l'       & \alpha_3 \\
                       s' & {\alpha_2} & j'
                   \end{array}} \right \}
       (-1)^{f_1} \, C(f_1 j' \alpha_2 ; 0 {\mu}' {\mu}') 
\]
\[  \times
\sum\limits_{f_2} \sqrt{\hat{f_2}} 
                 \left \{ { \begin{array}{ccc}
                       f_1  &    f_2   & e \\
                       l & {\alpha_3} & l'
                   \end{array}} \right \} \
\sqrt{ ( 2 f_2 + 1 ) ! } 
\]
\[  \times
\sum\limits_{r} \hat{r} 
C( r d_2 f_1 ; 0 0 0 )
C( r c f_2  ; 0 0 0 )
                 \left \{ { \begin{array}{ccc}
                       d_2  &    c   & e \\
                       f_2 & f_1 & r
                   \end{array}} \right \} \
\]
\[  \times
\sum\limits_{u_1 + u_2 = f_2} (p')^{\, u_1} (p)^{u_2} 
   \frac{1}{ \sqrt{ ( 2 u_1 + 1 ) !  ( 2 u_2 ) ! }}
\]
\begin{equation}
\times
\sum\limits_{z} \sqrt{\hat{z}} C( u_2 l z ; 0 0 0 ) C( l' z u_1 ; 0 0 0 ) 
                 \left \{ { \begin{array}{ccc}
                       u_1  &    u_2   & f_2 \\
                       l & l' & z
                   \end{array}} \right \} \ H^{h_1 w_1 f_2}_{z r} {\rm ,} 
\label{jrhoIII}
\end{equation}
where 
\begin{equation}
 H^{h_1 w_1 f_2}_{z r} = -  \int\limits_{-1}^1 d y P_z (y) 
                  \Delta^{ 1 + h_1 + w_1 - f_2 } \, 
                     \int\limits_{-1}^1 d x P_r (x) 
  \frac { v_\rho ( \frac14 Q^2 + \Delta^2 - Q \Delta x )
       -  v_\rho ( \frac14 Q^2 + \Delta^2 + Q \Delta x )}
        { 2 Q \Delta x } {\rm .}
\end{equation}

\[
\tilde I_2 ^{IV \rho} ( p',p,Q; (l's')j' \mu ', (ls) j \mu )
\]

\[ = -36 \pi \sqrt{3}
     \sqrt{ \hat{l} \hat{s} \hat{j} \hat{l'} \hat{s'} \hat{j'} }
(-1)^{l+s+j} \, \delta_{{\mu}',{\mu}+\zeta } \ 
\]
\[  \times
\sum\limits_{\alpha_1} \hat{{\alpha_1}} 
  \ \left \{ { \begin{array}{ccc}
                       1  &1 &  {\alpha_1}  \\
                   \frac12   & \frac12 & s  \\
                    \frac12  & \frac12 & s'
                   \end{array}} \right \}
\]
\[  \times
\sum\limits_{\alpha_2}   \hat{{\alpha_2}}
  \ \left \{ { \begin{array}{ccc}
                       1  &1 &  1  \\
                       \alpha_2 & \alpha_1 & 1
                   \end{array}} \right \}
\]
\[  \times
\sum\limits_{u_1 + u_2 = 1} ( \frac12 Q )^{u_2} 
\]
\[  \times
\sum\limits_{\alpha_3} \, C( 1 j {\alpha_3} ; \zeta  \mu {\mu}' )
\]
\[  \times
\sum\limits_{h}   \hat{h}
  \ \left \{ { \begin{array}{ccc}
                       \alpha_1 & \alpha_2 & 1 \\
                       s  & l &  j  \\
                       s' & h &  \alpha_3
                   \end{array}} \right \}
\]
\[  \times
\sum\limits_{f_1} \sqrt{\hat{f_1}} C( 1 u_1 f_1 ; 0 0 0 ) 
\]
\[  \times
\sum\limits_{f_2} \sqrt{\hat{f_2}} C( 1 u_2 f_2 ; 0 0 0 ) 
                 \left \{ { \begin{array}{ccc}
                       1  &  1 &  1 \\
                       u_1  &   u_2  & 1 \\
                       f_1  &   f_2  & \alpha_1
                   \end{array}} \right \}
\]
\[  \times
\sum\limits_{g_1} \sqrt{\hat{g_1}} \sqrt{ ( 2 g_1 + 1 ) ! } 
\]
\[  \times
\sum\limits_{g_2} \sqrt{\hat{g_2}} 
                \left \{ { \begin{array}{ccc}
                       g_2 & l' & h \\
                       s' & \alpha_3 & j'
                   \end{array}} \right \}
               \  \left \{ { \begin{array}{ccc}
                       g_2 & g_1 & \alpha_2 \\
                       l & h & l'
                   \end{array}} \right \}
\]
\[  \times
\sum\limits_{r} \hat{r} (-1)^r \
C( r f_1 g_1 ; 0 0 0 )
C( r f_2 g_2  ; 0 0 0 )
                 \left \{ { \begin{array}{ccc}
                       g_2  &  g_1  & \alpha_2 \\
                       f_1 & f_2 & r
                   \end{array}} \right \} \
  [ 1 + (-1)^{\alpha_1 + u_1 + r}]
\]
\[  \times
\sum\limits_{w_1 + w_2 = g_1} (p')^{\, w_1} (p)^{w_2}  (-1)^{w_2}
   \frac{1}{ \sqrt{ ( 2 w_1 + 1 ) !  ( 2 w_2 ) ! }}
\]
\begin{equation}
\times
\sum\limits_{z} \sqrt{\hat{z}} (-1)^{z} 
      C( w_2 l z ; 0 0 0 ) C( l' z w_1 ; 0 0 0 ) 
                 \left \{ { \begin{array}{ccc}
                       w_1  &    w_2   & g_1 \\
                       l & l' & z
                   \end{array}} \right \} \ H^{u_1 g_1 }_{z r} {\rm ,} 
\label{jrhoIV}
\end{equation}
where 
\begin{equation}
 H^{u_1 g_1}_{z r} = -  \int\limits_{-1}^1 d y P_z (y) 
                  \Delta^{ 1 + u_1 - g_1 } \, 
                     \int\limits_{-1}^1 d x P_r (x) 
  \frac { v_\rho ( \frac14 Q^2 + \Delta^2 - Q \Delta x )
       -  v_\rho ( \frac14 Q^2 + \Delta^2 + Q \Delta x )}
        { 2 Q \Delta x } {\rm .}
\end{equation}

Secondly we present the form required for the 4-fold integration. 
The separation of spin and momentum parts like in (\ref{MECSpIsoSp}),
is also displayed 
for the convenience of the reader. The first term $j^{I,\rho}$ has
no spin-dependence and is left at it is.


\be
\tilde I_2 ^{II \rho} ( p',p , Q; (l's')j' \mu ', (ls) j \mu )
\ee 
\[
= \sum _{m,m'} C (l s j , \mu -m , m ) C (l' s' j' , \mu ' -m' , m' )
\int d \hat p d\hat {p'} Y^* _{l' \mu'-m' } (\hat {p'})
 Y_ {l \mu -m } (\hat p ) 
\]
\[
\times 6 \sum _{\alpha_1 \nu}
 \left \{ { \begin{array}{ccc}
           1  &    1   & \alpha_1  \\
              1 & 1 & 1
                   \end{array}} \right \}
C( \alpha_1 1 1, \nu-\zeta, \zeta, \nu ) 
\left\{ \sigma(2) \otimes \sigma(3) \right\}_{\alpha_1 \nu}
\]
\[ \times
[ -v_\rho (p_3)(-)^{-\nu+\zeta}(p_3)_{\zeta-\nu} + v_\rho (p_2)
(-)^{\alpha_1 -\nu+\zeta} (p_2)_{\zeta-\nu} ]  {\rm ,}
\]

\be
\tilde I_2 ^{III \rho} ( p',p , Q; (l's')j' \mu ', (ls) j \mu )
\ee 
\[
= \sum _{m,m'} C (l s j , \mu -m , m ) C (l' s' j' , \mu ' -m' , m' )
\int d \hat p d\hat {p'} Y^* _{l' \mu'-m' } (\hat {p'})
 Y_ {l \mu -m } (\hat p ) 
\]
\[ \times
\frac{ ( {\vec p}_2 - {\vec p}_3 )_{\zeta} }{ p_2^2 - p_3^2 } \
\left( v_\rho (p_2) - v_\rho (p_3) \right)
\]
\[
\times 
6 \sum_{\alpha_1 \nu} (-) ^{ 1 - \nu }  
 \left \{ { \begin{array}{ccc}
           1  &    1   & 1  \\
              1 & 1 & \alpha_1 
                   \end{array}} \right \}
\left\{ \sigma(2) \otimes \sigma(3) \right\}_{\alpha_1 \nu}
\left\{ p_2  \otimes p_3 \right\}_{\alpha_1, -\nu} {\rm ,}
\]

\be
\tilde I_2 ^{IV \rho} ( p',p , Q; (l's')j' \mu ', (ls) j \mu )
\ee 
\[
= \sum _{m,m'} C (l s j , \mu -m , m ) C (l' s' j' , \mu ' -m' , m' )
\int d \hat p d\hat {p'} Y^* _{l' \mu'-m' } (\hat {p'})
 Y_ {l \mu -m } (\hat p ) 
\]
\[ \times
\frac{   v_\rho (p_2) - v_\rho (p_3)}{ p_2^2 - p_3^2 } \
\sum_{\alpha_1 \alpha_2} (-2) \sqrt{ \hat \alpha_1 \hat \alpha_2} 
 \left \{ { \begin{array}{ccc}
           1  &    1   & 1  \\
              \alpha_1 & \alpha_2  & 1 
                   \end{array}} \right \}
(-1)^{\alpha_2}
\]
\[
\times
\sum_{\kappa} 
C(\alpha_1 \alpha_2 1 , \kappa, \zeta - \kappa, \zeta ) 
\left\{ \sigma(2) \otimes \sigma(3) \right\}_{\alpha_1 \kappa}
\sum_{\kappa_2 \kappa_3}
C( 1 1 \alpha_2, -\kappa_2-\kappa_3+\zeta-\kappa, \kappa_2+\kappa_3, \zeta -\kappa)
\]
\[
\times
C( 1 1 1, \kappa_2 , \kappa_3, \kappa_2 + \kappa_3 ) 
\left\{ (p_2)_{\kappa_4} (p_2)_{\kappa_2} (p_3)_{\kappa_3} 
+ (-)^{1 + \alpha_1} (p_3)_{\kappa_4} (p_3)_{\kappa_2} (p_2)_{\kappa_3}
\right\} {\rm ,}
\]
with $\kappa_4 = -\kappa_2 - \kappa_3 +\zeta -\kappa $.
The momenta $p_2$ and $p_3$ are given in  Eqs. (\ref{p23}).

\setcounter{section}3
\section{Functions $F$ and $G$ in the Case of $\pi$--Meson Exchange Currents}
\label{TensorsFG}
\setcounter{equation}0
We have for the seagull contributions (\ref{Glm})--(\ref{Glmu})
\be\label{GlmPV}
\begin{array}{lll}\lefteqn{
G^{l'm'}_{lm}(n) =
\delta_{m',m}
} &&\\&\times&
{\dss {1\over\pi}}
{\dss\int\limits_{-1}^{+1} dx \int\limits_{-1}^{+1} dx'} \:
\bar{P}^{m'}_{l'}(x')  \bar{P}^m_l(x)
V(m,n),
\end{array}\ee
\be\label{GlmuPV}
\begin{array}{lll}\lefteqn{
G^{l'm'}_{lm\mu}(n) =
\delta_{m',m+\mu}
} &&\\&\times&
{\dss {1\over\pi}}{\dss\left({2\over3}\right)^{1/2}}
{\dss\int\limits_{-1}^{+1} dx \int\limits_{-1}^{+1} dx'} \:
\bar{P}^{m'}_{l'}(x')  \bar{P}^m_l(x)
\\&\times&
\left(p'\bar{P}^{\mu}_1(x')V(m,n)-
      p\bar{P}^{\mu}_1(x)V(m+\mu,n)\right)
\end{array}\ee
and for the ones (\ref{Fmu1})--(\ref{Fmu3}) of the pionic current
\be\label{Fmu1PW}
\begin{array}{lll}\lefteqn{
F^{l'm'}_{lm\mu} =
\delta_{m',m+\mu}
} &&\\&\times&
{\dss {1\over\pi}}{\dss\left({2\over3}\right)^{1/2}}
{\dss\int\limits_{-1}^{+1} dx \int\limits_{-1}^{+1} dx'} \:
\bar{P}^{m'}_{l'}(x')  \bar{P}^m_l(x)
\\&\times&
\left(p'\bar{P}^{\mu}_1(x')W(m)-
      p \bar{P}^{\mu}_1(x) W(m+\mu)\right),
\end{array}\ee
\be\label{Fmu2PW}
\begin{array}{lll}\lefteqn{
F^{l'm'}_{lm\mu_1\mu_2} =
\delta_{m',m+\mu_1+\mu_2}
} &&\\&\times&
{\dss {2\over3\pi}}
{\dss\int\limits_{-1}^{+1} dx \int\limits_{-1}^{+1} dx'} \:
\bar{P}^{m'}_{l'}(x')  \bar{P}^m_l(x)
\\&\times&
\left(
p'^2\bar{P}^{\mu_1}_1(x')\bar{P}^{\mu_2}_1(x') W(m)
\right.
\\&-&
      pp' \bar{P}^{\mu_1}_1(x')\bar{P}^{\mu_2}_1(x ) W(m+\mu_2)
\\&-&
      pp' \bar{P}^{\mu_1}_1(x )\bar{P}^{\mu_2}_1(x') W(m+\mu_1)
\\&+&
\left.
      p^2 \bar{P}^{\mu_1}_1(x')\bar{P}^{\mu_2}_1(x ) W(m+\mu_1+\mu_2)
\right),
\end{array}\ee
\be\label{Fmu3PW}
\begin{array}{lll}\lefteqn{
F^{l'm'}_{lm\mu_1\mu_2\mu_3}=
\delta_{m',m+\mu_1+\mu_2+\mu_3}
} &&\\&\times&
{\dss {1\over\pi}}{\dss\left({2\over3}\right)^{1/2}}
{\dss\int\limits_{-1}^{+1} dx \int\limits_{-1}^{+1} dx'} \:
\bar{P}^{m'}_{l'}(x')  \bar{P}^m_l(x)
\\&\times&
\left( 
p'^3   \bar{P}^{\mu_1}_1(x')\bar{P}^{\mu_2}_1(x')\bar{P}^{\mu_3}_1(x')
W(m)
\right.
\\&-&  
p'^2p  \bar{P}^{\mu_1}_1(x )\bar{P}^{\mu_2}_1(x')\bar{P}^{\mu_3}_1(x')
W(m+\mu_1)
\\&-&  
p'^2p  \bar{P}^{\mu_1}_1(x')\bar{P}^{\mu_2}_1(x )\bar{P}^{\mu_3}_1(x')
W(m+\mu_2)
\\&+&  
p'p^2  \bar{P}^{\mu_1}_1(x )\bar{P}^{\mu_2}_1(x )\bar{P}^{\mu_3}_1(x')
W(m+\mu_1+\mu_2)
\\&-&  
p'^2p  \bar{P}^{\mu_1}_1(x')\bar{P}^{\mu_2}_1(x')\bar{P}^{\mu_3}_1(x )
W(m+\mu_3)
\\&+&  
p'p^2  \bar{P}^{\mu_1}_1(x )\bar{P}^{\mu_2}_1(x')\bar{P}^{\mu_3}_1(x )
W(m+\mu_1+\mu_3)
\\&+&  
p'p^2  \bar{P}^{\mu_1}_1(x')\bar{P}^{\mu_2}_1(x )\bar{P}^{\mu_3}_1(x )
W(m+\mu_2+\mu_3)
\\&-&  
\left.
p'^3   \bar{P}^{\mu_1}_1(x )\bar{P}^{\mu_2}_1(x )\bar{P}^{\mu_3}_1(x )
W(m+\mu_1+\mu_2+\mu_3)
\right)
\end{array}\ee
with
\be
\bar{P}^{m}_l (cos\theta) \equiv 
\sqrt{2 \pi}~~e^{-im\phi}~~Y_{lm}(\theta, \phi).
\ee

Note, that formulae (\ref{GlmPV})--(\ref{Fmu3PW}) have been obtained in the coordinate system
where $\vQ\parallel{\vec e}_z$.

\newpage

\section{Integrals over the Azimuthal Angle}
\label{TableInt}
\setcounter{equation}0

The integrands of (\ref{Glm})--(\ref{Fmu3}) contain functions
$$
v(p)=\Cpi\Fpi(\mpi+p^2)^{-1}
\qquad \mbox{and} \qquad
H(p_2,p_3)=(p_3^2-p_2^2)^{-1}(v(p_2)-v(p_3)).
$$

In Appendix D we list the expressions for $F$ and $G$
(Eqs.(\ref{Glm})--(\ref{Fmu3})) in terms of two--dimensional integrals over
the products of the associated Legendre polynomials ${\bar P}^m_l(x)$
and the integrals
\be\label{Vmnt}\label{Wmt}
\begin{array}{rcl}
V(m,n)= {\dss\int\limits_{0}^{2\pi}} dt(2t-t)\cos(mt)v(p_n),
&\mbox{and}&
W(m)  = {\dss\int\limits_{0}^{2\pi}} dt(2t-t)\cos(mt)H(p_2,p_3).
\end{array}
\ee
Functions (\ref{Vmnt})
belong to the type of integral (\ref{RMCos}).
To compute (\ref{Vmnt}) analytically with $v(p_n)$ and $H(p_2,p_3)$
in question we write for the vector $\vec p_n$ (\ref{vecpn})
\be
p_n^2=a_n+b\cos t
\ee
where
\bea
a_n&=&{\dss 1\over4} Q^2+p'^2+p^2+(-1)^n\vQ(\vpp-\vp)-2p'p\:\cos\theta\:\cos\theta',
\\
b&=&-2p'p\:\sin\theta\:\sin\theta'
\eea
and decompose
\bea
v(p_n)&=&\Cpi\left[(a_n+\mpi+b\cos t)^{-1}-(a_n+\Lpi+b\cos t)^{-1}
\right.
\nonumber\\&&\phantom{\Cpi}\left.
-(\Lpi-\mpi)(a_n+\Lpi+b\cos t)^{-2}\right],
\\
H(p_2,p_3)&=&\Cpi(a_2-a_3)^{-1}                                
\nonumber\\&& \left[
-(a_2+\mpi+b\cos t)^{-1}+(a_3+\mpi+b\cos t)^{-1} \right.
\nonumber\\&&
+(a_2+\Lpi+b\cos t)^{-1}-(a_3+\Lpi+b\cos t)^{-1}
\nonumber\\&& \left.
+(\Lpi-\mpi)((a_2+\Lpi+b\cos t)^{-2}-(a_3+\Lpi+b\cos t)^{-2} \right].
\eea

We get
\bea
V(m,n)&=&                                                      
\Cpi \left[V(a_n+\mpi,\:b,\:m,\:1)-V(a_n+\Lpi,\:b,\:m,\:1)\right.
\nonumber\\&&\phantom{\Cpi}\left.
-(\Lpi-\mpi)V(a_n+\Lpi,\:b,\:m,\:2)\right],
\\
W(m)&=&\Cpi(a_2-a_3)^{-1}                                      
\nonumber\\&&\phantom{\Cpi}\left[
-V(a_2+\mpi,\:b,\:m,\:1)+V(a_3+\mpi,\:b,\:m,\:1) \right.
\nonumber\\&&\phantom{\Cpi}
+V(a_2+\Lpi,\:b,\:m,\:1)-V(a_3+\Lpi,\:b,\:m,\:1)
\nonumber\\&&\phantom{\Cpi}\left.
+(\Lpi-\mpi)(
 V(a_2+\Lpi,\:b,\:m,\:2)-V(a_3+\Lpi,\:b,\:m,\:2))\right].
\eea

Integrals (\ref{VabmnInt})
can be calculated in closed form, e.g., with the help of tables \cite{GradshteynRyzhik}.
For $m\ge 0$ and $n=1,2$ the result reads
\bea\label{VabmnClosedF}
V(a,b,m,n)&=&2\pi^2
\left[{\dss1\over b} \left(\sqrt{a^2-b^2}-a\right)\right]^m
\nonumber\\
&\times&
(a^2-b^2)^{-n+\oh}\:
\left(a+m\sqrt{a^2-b^2}\right)^{n-1},
\eea
for $m,0$ one can use the relation $V(m,n)=V(-m,n).$


\vfil \eject

\end{document}